\documentclass[twocolumn]{aastex631}

\usepackage{tabularx}
\usepackage{newtxtext,newtxmath}
\usepackage[T1]{fontenc}
\usepackage{eucal}
\usepackage{nicefrac}
\usepackage{amssymb}
\usepackage{latexsym}
\usepackage{amsmath, mathrsfs, bm}
\usepackage{eqnarray}
\usepackage{url}
\usepackage{cases}
\usepackage{epsfig}
\usepackage{color}
\usepackage{graphicx}
\usepackage{grffile}
\usepackage{float}
\usepackage{soul}
\usepackage{enumerate}
\usepackage{multirow}
\usepackage{threeparttable}
\usepackage{xspace}
\usepackage{bm}
\usepackage{times}
\usepackage{etoolbox}
\usepackage{multirow}

\usepackage{xcolor}
\definecolor{green}{RGB}{50,205,50}

\begin{document}

\title{Exploring the Key Features of Repeating Fast Radio Bursts with Machine Learning }

\author[0000-0001-5950-7170]{Wan-Peng Sun}
\affiliation{Liaoning Key Laboratory of Cosmology and Astrophysics, College of Sciences, Northeastern University, Shenyang 110819, China}

\author[0009-0004-5607-9181]{Ji-Guo Zhang}
\affiliation{Liaoning Key Laboratory of Cosmology and Astrophysics, College of Sciences, Northeastern University, Shenyang 110819, China}

\author[0000-0003-1962-2013]{Yichao Li}
\affiliation{Liaoning Key Laboratory of Cosmology and Astrophysics, College of Sciences, Northeastern University, Shenyang 110819, China}

\author[0009-0004-8194-7446]{Wan-Ting Hou}
\affiliation{College of Mathematics and Statistics, Liaoning University, Shenyang 110036, China}

\author[0000-0002-5936-8921]{Fu-Wen Zhang}
\affiliation{College of Science, Guilin University of Technology, Guilin 541004, China}

\author[0000-0002-3512-2804]{Jing-Fei Zhang}
\affiliation{Liaoning Key Laboratory of Cosmology and Astrophysics, College of Sciences, Northeastern University, Shenyang 110819, China}

\author[0000-0002-6029-1933]{Xin Zhang}
\affiliation{Liaoning Key Laboratory of Cosmology and Astrophysics, College of Sciences, Northeastern University, Shenyang 110819, China}
\affiliation{MOE Key Laboratory of Data Analytics and Optimization for Smart Industry, Northeastern University, Shenyang 110819, China}
\affiliation{National Frontiers Science Center for Industrial Intelligence and Systems Optimization, Northeastern University, Shenyang 110819, China}

\correspondingauthor{Yichao Li}
\email{liyichao@mail.neu.edu.cn}

\correspondingauthor{Jing-Fei Zhang}
\email{jfzhang@mail.neu.edu.cn}

\correspondingauthor{Xin Zhang}
\email{zhangxin@mail.neu.edu.cn}

\begin{abstract}

Fast radio bursts (FRBs) are enigmatic high-energy events with unknown origins, which are observationally divided into two categories, i.e., repeaters and non-repeaters. However, there are potentially a number of non-repeaters that may be misclassified, as repeating bursts are missed due to the limited sensitivity and observation periods, thus misleading the investigation of their physical properties. In this work, we propose a repeater identification method based on the t-distributed Stochastic Neighbor Embedding algorithm and apply the classification to the 
first Canadian Hydrogen Intensity Mapping Experiment Fast Radio Burst (CHIME/FRB) catalog. We find that the spectral morphology parameters, specifically spectral running ($r$), represent the key features for identifying repeaters from the non-repeaters. Also, the results suggest that repeaters are more biased toward narrowband emission, whereas non-repeaters are inclined toward broadband emission. We provide a list of 163 repeater candidates, five of which are confirmed with an updated repeater catalog from CHIME/FRB. Our findings improve our understanding of the various properties underlying repeaters and non-repeaters, as well as guidelines for future FRB detection and categorization.

\end{abstract}


\section{Introduction} \label{sec:intro}

Fast radio bursts (FRBs) are a class of radio flashes with violent explosion energies and millisecond durations. 
The dispersion measures (DMs) of the FRBs are beyond the maximum contribution of the Milky Way, except for one FRB, which has been found to be associated with a magnetar located in the Milky Way \citep{2020Natur.587...59B}. 
FRBs have received extensive attention since their discovery in 2007 \citep{2007Sci...318..777L,2013Sci...341...53T,2022A&ARv..30....2P,2023RvMP...95c5005Z}.
Although the physical origin of FRBs is unknown, their unique properties and observations can be utilized to probe fundamental physics, 
astrophysics, and cosmology 
\citep{2014ApJ...783L..35D,2020Natur.581..391M,2020ApJ...903...83Z,2022JCAP...02..006Q,2022ApJ...940L..29Y,2023ApJ...944...50W,2023ApJ...955..101W,2023JCAP...06..052W,2023SCPMA..6620412Z,2023JCAP...04..022Z}. 
Numerous experiments, including the Canadian Hydrogen Intensity Mapping Experiment (CHIME) \citep{2018ApJ...863...48C}, 
Five-hundred-meter Aperture Spherical radio Telescope \citep{2016RaSc...51.1060L}, Tianlai Pathfinder \citep{2024RAA....24h5010Y}, and Australian Square Kilometer Array Pathfinder \citep{2024arXiv240802083S} 
are currently engaged in FRB research. Their efficient observational capabilities have 
significantly enriched the FRB sample. 
So far, over 800 named FRBs have been reported from over 600 sources\footnote{\url{https://blinkverse.alkaidos.cn}}, 
with more than 200 repeating FRBs from 25 sources and 43 host galaxies located among all observed FRBs \citep{2024MNRAS.528.6340P}. 

Recent observational evidence has significantly advanced theoretical frameworks for the origin and energy mechanisms of FRBs. 
The extremely high brightness temperature, e.g. with a maximum of $\sim 10^{36}$ K, indicates a possible coherent radiation mechanism
\citep{2014PhRvD..89j3009K,2014ApJ...785L..26L,2019A&ARv..27....4P,2021ApJ...922..166L,2024ApJ...972..124Q}.
The millisecond-order of magnitude duration ($20~{\rm \mu s} - 30~{\rm ms}$) gives an upper limit to the 
emitting region of FRBs at $\lesssim 10~{\rm km}$ (without the presence of relativistic beaming effects) 
and a corresponding possible model of origin \citep{2018Natur.553..182M}. Despite significant advances in energy mechanisms and theoretical frameworks \citep{2019A&ARv..27....4P,2019ARA&A..57..417C,2020Natur.587...45Z,
2021SCPMA..6449501X,2022A&ARv..30....2P,2023RvMP...95c5005Z}, multi-wavelength observations and the localization of high-energy counterparts remain challenging \citep{2021Univ....7...76N}. While repeating FRBs can originate from magnetar models \citep{2020Natur.587...54C,2020ApJ...898L..29M,2020Natur.587...59B,2021NatAs...5..378L,2021NatAs...5..372R,2021NatAs...5..401T}, catastrophic models for non-repeating FRBs, such as binary neutron star mergers or massive neutron star collapses \citep{2014MNRAS.441.2433R,2017arXiv170102492D}, still lack definitive observational evidence.

The FRBs can be divided into repeating and non-repeating samples. However, there is a significant disparity in the number of observations between them. In the published observational samples, non-repeating FRBs account for 91.88\%, 
while repeating FRBs only account for 8.12\% \citep{2023Univ....9..330X}. Recent studies suggest that most observed FRBs may originate from repeating sources, with the intrinsic proportion of repeaters potentially exceeding 50\%. \citep{2019NatAs...3..928R, 2024MNRAS.52711158Y, 2024arXiv241017474H}. This is completely inconsistent with the current observation \citep{2023ApJ...947...83C}. With sufficient follow-up observations, some sources initially thought to be non-repeating have been confirmed as repeating \citep{2022A&ARv..30....2P,2023ApJ...947...83C}. Therefore, based on these findings, it is plausible that due to uncertainties in observational periods, equipment sensitivity limitations, and the low brightness of some bursts, many subsequent FRBs from repeating sources may be missed and misclassified as non-repeaters. 

Hence, relying solely on observationally identified repeaters and non-repeaters is incomplete. It may be possible to further distinguish which are repeaters and non-repeaters based on their observational features.
Generally, compared to non-repeating FRBs, repeating FRBs have wider pulse widths 
\citep{2016ApJ...833..177S,2021ApJS..257...59C,2021ApJ...923....1P,2022ApJ...926..206Z}, 
narrower frequency bandwidths \citep{2021ApJS..257...59C}, 
larger positive spectral index \citep{2021ApJ...923....1P,2022ApJ...926..206Z}, 
more common downward frequency drifts \citep{2019ApJ...885L..24C}, 
and lower peak luminosities and energies \citep{2022ApJ...926..206Z}. 
However, characterizing the repeating FRBs according to the single observation feature is not straightforward. 
All the FRB observational features form a high-dimensional parameter space and the 
repeating FRB identification is equivalent to a high-dimensional classification problem.

Machine learning offers a unique approach to addressing this issue and has been explored based on unsupervised \citep{2022MNRAS.509.1227C,2023MNRAS.522.4342Y,2023MNRAS.519.1823Z,2024arXiv241102216G,2024arXiv241114040Q}
and supervised machine learning techniques \citep{2023MNRAS.518.1629L,2024MNRAS.533.3283S}. Recent observations indicate a significant variance in the reliability of these methods \citep{2023ApJ...947...83C}.
Indeed, selecting the appropriate features to input into the machine learning algorithm is paramount. 
More features provide richer information but can dilute the most critical ones. Irrelevant features can obscure the physical differences between repeaters and non-repeaters. 
Hence, it is important to identify the most critical and effective observational parameters for machine learning to distinguish between them. We aim to determine which observational features best reflect the differences between repeaters and non-repeaters.

In this work, we employ an unsupervised machine learning algorithm, t-distributed Stochastic Neighbour Embedding (t-SNE), with feature preselection to identify the potential repeaters in the first CHIME/FRB catalog. The structure of our paper is as follows: In Section \ref{sec:data}, we introduce the selected CHIME/FRB data sample, the t-SNE algorithm and corresponding parameter configuration. Section \ref{sec:resu} presents the categorization findings of repeaters and non-repeaters found using t-SNE, as well as a detailed comparison of the statistical differences between the two types of FRBs. In Section \ref{sec:concl}, we summarize our work. The cosmological parameters in this paper are $H_{0}=67.66$ km s$^{-1}$ Mpc$^{-1}$, $\Omega_{\rm{b}}=0.04897$, $\Omega_{\rm{m}} = 0.3111$, and $\Omega_{\Lambda}=0.6874$ from {\tt Planck2018} \citep{2020A&A...641A...6P}.

\section{DATA AND ALGORITHM PARAMETERS} \label{sec:data}

\subsection{Data Sample Selection} \label{subsec:data sel}

We use 536 FRBs published with the first CHIME/FRB catalog, which includes 474 non-repeaters and 62 repeaters. 
The FRBs in the first CHIME/FRB catalog were detected with observations from 2018 July 25 to 2019 July 1. 
These observations were carried out with a frequency range of $400$ MHz to $800$ MHz \citep{2021ApJS..257...59C}. 
We exclude six non-repeaters that have no flux or fluence observation data. 
In addition, some FRBs contain sub-bursts, where a sub-burst refers to a notable local peak in intensity within the profile of an FRB, particularly when this structure appears distinct in frequency from the surrounding emission in the dynamic spectrum, similar to the definition by \citet{2024MNRAS.52710425S}. The FRBs' sub-bursts are considered as individual bursts in this analysis.
So, we compile a sample of 500 non-repeaters and 94 repeaters as original sample (\textbf{S\textsubscript{org}}). 

\subsection{FRB Feature Parameters} \label{ssec:data desc}

\begin{table*}\scriptsize
\begin{center}
\caption{FRB feature parameters of the first CHIME/FRB catalog.}
\label{tab:tab1}
\begin{tabularx}{\textwidth}{l l X}
\hline
\bm{}  Parameter  &  Unit  &  Description  \\
\hline
$S_{\nu}$   &  (Jy)  &  Peak flux of band-average profile.The value is the same for all sub-bursts from an FRB.  \\

$F_{\nu}$  &  (Jy ms)  &  This value represents the apparent brightness integrated over of all sub-bursts. These values are the same for each FRB sub-burst.  \\

$\Delta t_\mathrm{WS}$  &  (s)  &  These values represent the width of each sub-burst fitted by the fitBurst algorithm \citep{2015Natur.528..523M}.  \\

$\Delta t_\mathrm{ST}$  &  (s)  &  This value represents the scattering time of the burst at 600 MHz, which is the same between each sub-burst.  \\

$\gamma$  &    &  This value represents the spectral shape of every sub-burst. Between each sub-burst from an FRB, the spectral index is different.  \\

$r$  &    &  This value depicts the frequency dependency of the spectral shape. Spectral running is different from one sub-burst to another.  \\

$\nu_\mathrm{Low}$  &  (MHz)  &  These values represent the lowest frequency band of pulse detection at a full width tenth maximum. The value between each sub-burst from the FRB is different.  \\
\hline
$\Delta t_\mathrm{BW}$  &  (s)  &  Boxcar width means the duration of all sub-pulses combined, including instrumentation, scattering, and redshift broadening effects. The value is the same for each sub-burst of the FRB.  \\

$\nu_\mathrm{High}$  &  (MHz)  &  These values represent the highest frequency band of pulse detection at a full-width-tenth-maximum. The value between each sub-burst from the FRB is different.  \\

$\nu_\mathrm{Peak}$  &  (MHz)  &  These values represent the peak frequency of each sub-burst. The value between each sub-burst from the FRB is different.  \\

DM  &  (pc cm$^{-3}$)  &  These values represent observed measurements of dispersion. These values are the same for each sub-burst.  \\
\hline
\end{tabularx}
\end{center}
\end{table*}

Table \ref{tab:tab1} lists the 11 FRB observational parameters identified by CHIME, including the event name, location, signal-to-noise ratio (SNR), and fundamental physical measurements.

We focus on the feature parameters related to the FRBs' intrinsic or environmental properties.
The first two features considered are flux ($S_{\nu}$) and fluence ($F_{\nu}$), which characterize the fundamental physical properties of FRBs \citep{2019MNRAS.487..491K}. To characterize the FRB pulse width, we select the width of the sub-burst ($\Delta t_\mathrm{WS}$) instead of the boxcar width. The boxcar width typically includes additional intra-channel dispersion smearing and scattering, which do not robustly represent the intrinsic pulse width \citep{2021ApJS..257...59C}. The scattering time ($\Delta t_\mathrm{ST}$) is included to highlight its role in characterizing the complexity of the FRB host galaxy or source environment \citep{2022ApJ...931...87O,2024arXiv241102249Y}. 
Additionally, spectral index ($\gamma$) and spectral running ($r$) characterize the complexity 
of FRB spectra \citep{2021ApJS..257...59C}. According to \citet{2023ApJ...956...67Y,2024ApJ...974..160K}, the frequency spectrum reflects both the intrinsic radiation mechanisms and the effects of propagation environments. Especially as variations in plasma conditions and inhomogeneities along the propagation path, such as plasma lensing, scintillation, and filamentation instability, significantly influence the spectrum, the spectral complexity consequently offers insights into the characteristics of the propagation environment.

Finally, it should be noted that we consider only the lowest frequency ($\nu_\mathrm{Low}$) for the frequency characteristics, 
rather than both the lowest and highest frequency. 
Observations suggest that the FRB spectrum could span from a few hundred MHz to $\sim 8$ GHz \citep{2018ApJ...863....2G,2021ApJ...911L...3P}.
The CHIME's frequency band, i.e. $400$--$800$ MHz, covers the lower end of the FRB spectrum.
Thus, the FRBs' highest frequencies are mostly truncated by CHIME's instrumental upper band limit, and 
including the highest frequency may limit the intrinsic frequency bandwidth.
However, the FRBs' lowest frequencies are mostly observed within the CHIME frequency band
and are more plausible for analyzing CHIME/FRB data.
To avoid introducing potential errors for machine learning algorithms, we consider 
only $\nu_\mathrm{Low}$ for characterizing the frequency differences between repeaters and non-repeaters.

\begin{figure}[ht!]
\centering
\includegraphics[angle=0,scale=0.5]{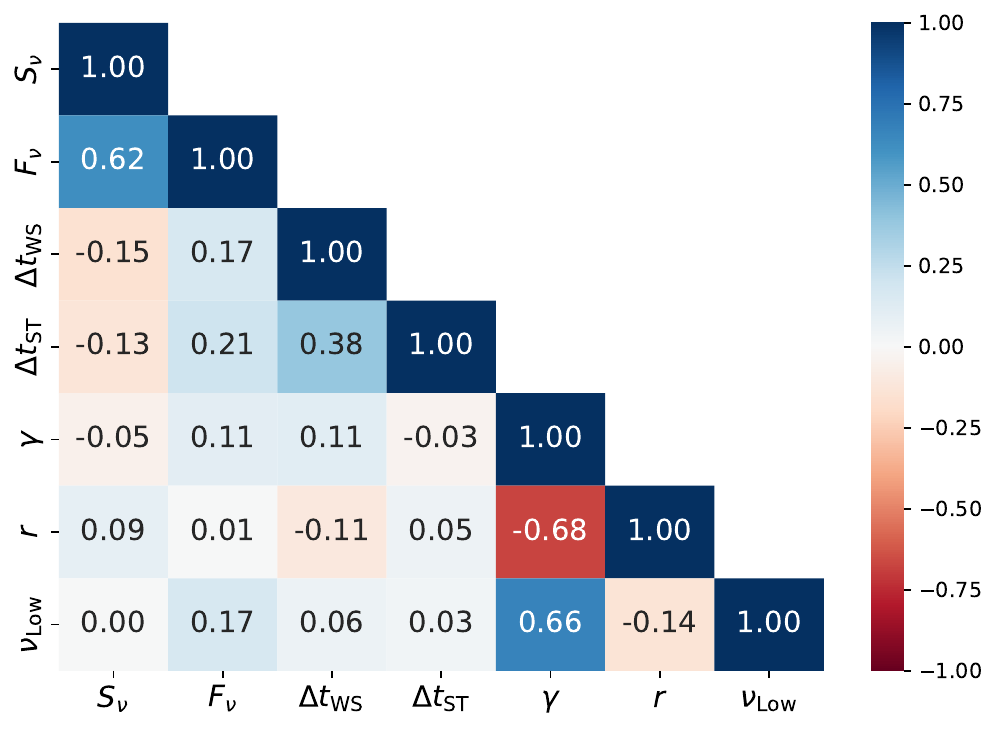}
\caption{The heatmap of Pearson correlation coefficient among the selected seven parameters. 
\label{fig:figure1}}
\end{figure}

In order to further ensure the rationality of the observational parameters selected for machine learning algorithms, 
we investigate the degeneracy among each parameter. As shown in Figure \ref{fig:figure1}, we plot the Pearson correlation coefficient
matrix of the selected seven parameters. There is a strong negative correlation between $\gamma$ and $r$, with a correlation coefficient of $-0.68$. Additionally, several pairs of observational parameters exhibit correlation coefficients greater than $0.5$.
Such apparent dependencies suggest a potential risk of multicollinearity among the parameters. 
To assess collinearity among the features, multicollinearity tests are conducted on the seven selected parameters. 

We adopt the variance inflation factor (VIF) to assess the multicollinearity and 
the VIF values are listed in Table \ref{tab:tab2}.
A higher VIF indicates severe multicollinearity issues. 
Typically, a VIF value exceeding 10 suggests unsuitability for statistical analysis. 
As shown in Table \ref{tab:tab2},
none of the selected seven parameters exceed this threshold. 

\begin{table}
\begin{center}
\caption{The variance inflation factor (VIF) from multicollinearity tests on the selected seven parameters.}
\label{tab:tab2}
\scriptsize
\setlength{\tabcolsep}{6pt}
\begin{tabular}{lccccccc}
\hline
Variable & $S_{\nu}$  &  $F_{\nu}$  &  $\Delta t_{WS}$  &  $\Delta t_{ST}$  &  $\gamma$  &  $r$  &  $\nu_{Low}$  \\
\hline
VIF    &  1.95  &  2.00  &  1.37  &  1.39  &  3.80  &  2.51  &  2.16  \\ 
\hline
\end{tabular}
\end{center}
\end{table}

\subsection{t-SNE Algorithm} \label{ssec:algor}

t-SNE is a machine learning algorithm for dimensionality reduction. It is commonly applied to project high-dimensional data into
the embedding space with two or three dimensions \citep{JMLR:v9:vandermaaten08a,JMLR:v15:vandermaaten14a}. 
The t-SNE algorithm is nonlinear and preserves local data structure by converting high-dimensional distances into conditional probabilities, which guide the arrangement of points in a low-dimensional embedding initialized randomly. This process helps t-SNE reveal clusters and patterns hidden in high-dimensional datasets.
Based on random initialization, the t-SNE method can generate different 
topology structures in the embedding space \citep{2020ApJ...891..136S}. 
However, in whatever topology structure of the embedding space, 
data points that were more alike in the original high-dimensional 
space remain closer together. 
It should be noted that the coordinate axes of the embedding space have no 
physical significance. They basically show that closer points are more similar
to one another, whereas distant points are less similar.
t-SNE has demonstrated robust performance and reliable results 
when used for dimensionality reduction in classifying gamma-ray bursts \citep{2020ApJ...896L..20J,2024MNRAS.527.4272C,2024MNRAS.532.1434Z}. 
It has also been utilized by \citet{2021ApJ...908..148H} to categorize Type 2 
active galactic nuclei and correct errors in galaxy properties. 

The key hyperparameters for the t-SNE initialization process include perplexity, early exaggeration, learning rate, and the number of steps. Perplexity determines the number of nearest neighbors considered when generating conditional probabilities. Larger datasets typically require higher perplexity values to capture more global structures, while with lower perplexity values the algorithm focuses on local structures. The t-SNE documentation suggests that perplexity values within the range of 5-100 generally yield meaningful results\footnote{\url{https://distill.pub/2016/misread-tsne/}}. For optimal outcomes, we recommend considering the stability of the recall value as an auxiliary criterion to determine the most suitable perplexity (see Section \ref{ssec:Stability} for details). In this work, perplexity is set to $63.0$ for optimization. Early exaggeration boosts the differences of 
high-dimensional data points during the initialization, which can help to
efficiently separate features in the lower-dimensional embedding. Varying the early exaggeration can significantly affect the topology structure in
the embedding space. However, it has little impact on similarity of data points. 
We use an early exaggeration value of $22.0$. 
Another crucial hyperparameter is the learning rate. 
A higher learning rate leads to overly uniform data visualization, where distances between points become nearly equal. Conversely, a lower learning rate tightly compresses data 
so that discerning the real structure becomes challenging.
In our work, the learning rate is fine-tuned to $265.0$. 
A detailed description of hyperparameters can be found in 
Scikit-learn\footnote{\url{https://scikit-learn.org/stable/modules/generated/sklearn.manifold.TSNE.html}}.

\section{RESULTS AND DISCUSSION} \label{sec:resu}

\subsection{t-SNE Results} \label{subsec:class}

\begin{figure}[ht!]
\centering
\includegraphics[angle=0,scale=0.28]{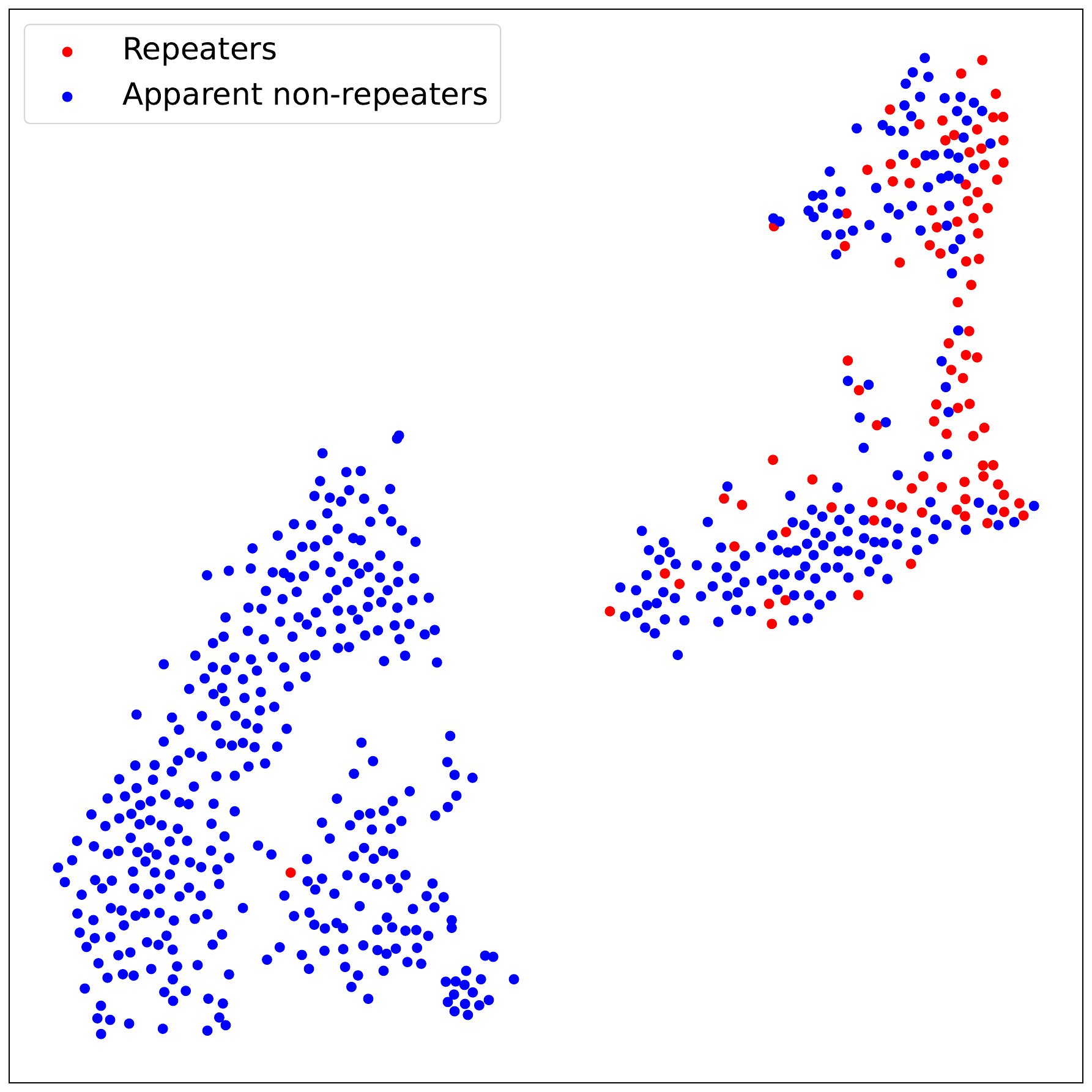}
\caption{
The embedding space of the t-SNE dimension reduction results of the selected 536 FRBs from the first CHIME/FRB catalog. 
The blue dots represent the apparent non-repeaters and the red dots are repeaters.
\label{fig:figure2}}
\end{figure}

The t-SNE dimension reduction results of the \textbf{S\textsubscript{org}} catalog are shown in Figure \ref{fig:figure2}.
Significantly, Figure \ref{fig:figure2} reveals a clear division of FRBs into two 
distinct clusters, with a pronounced separation between the left and right parts 
in Figure \ref{fig:figure2}. 
The red dots represent repeaters, while the green dots represent apparent non-repeaters. 
The result shows that almost all observed repeaters are gathered on the right side, 
while the left side is almost exclusively populated by observed apparent non-repeaters. 
As discussed in Section \ref{ssec:algor}, the coordinates of the t-SNE embedding space represent no physical meaning. However, in the embedding space,
FRBs close to each other have higher similarities.
A significant fraction of the apparent non-repeaters located on the right side
of Figure \ref{fig:figure2}, which are mixed with the observationally defined repeaters.
This suggests that such non-repeating FRBs could be candidates for repeating FRBs. 
The fact that these FRBs are currently classified as non-repeaters may be due to observational limitations.
This also implies a considerable distinction in the characteristics of repeaters 
and non-repeaters, which exerts a notable influence on unsupervised machine learning. 

\begin{figure}
\centering
\includegraphics[angle=0,scale=0.28]{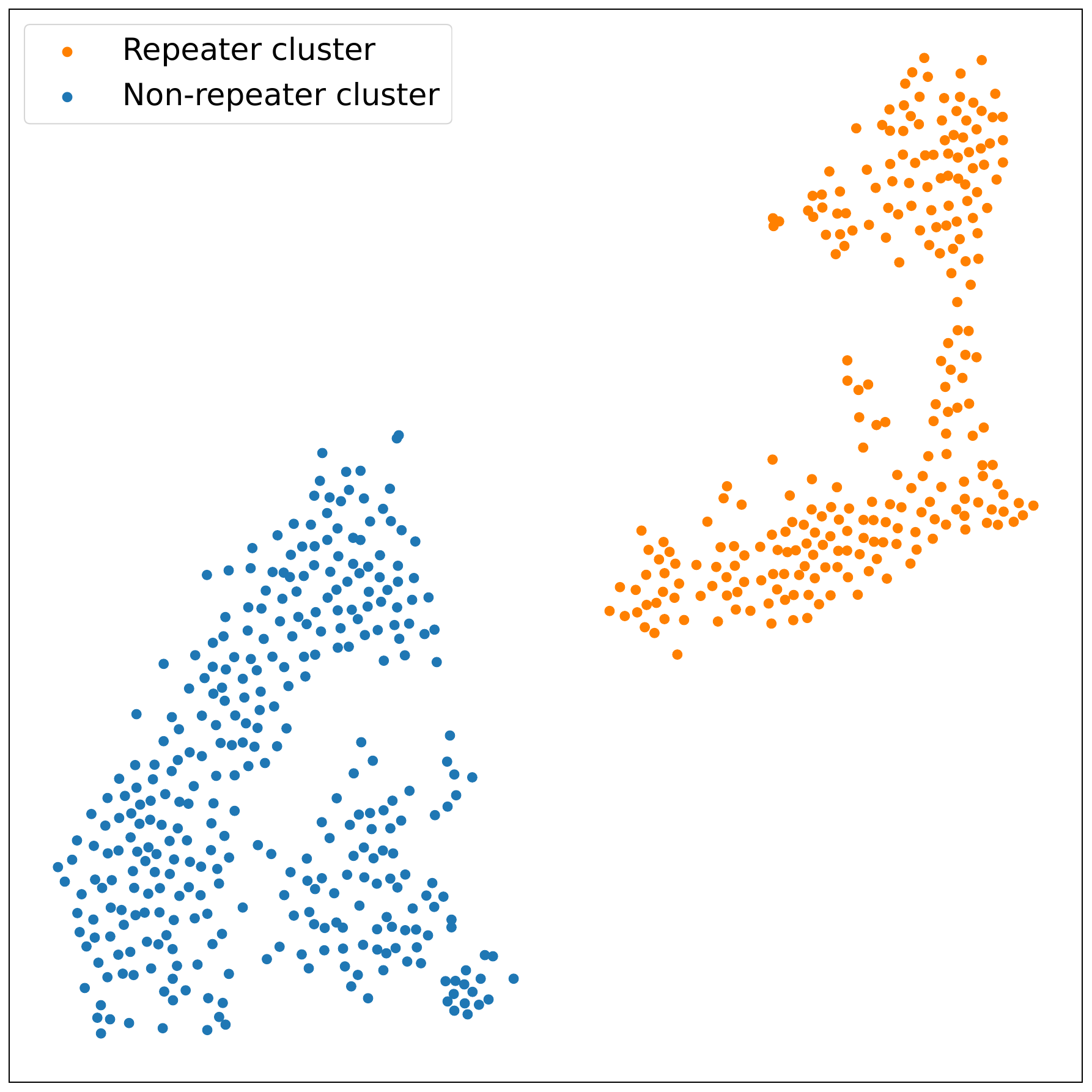}
\caption{The HDBSCAN clustering results of the FRB samples in the t-SNE embedding space. 
Non-repeaters are marked in blue, while repeaters are marked in orange.
\label{fig:figure3}}
\end{figure}

We further categorize the FRBs as repeaters and non-repeaters in the t-SNE embedding space
via the Hierarchical Density-Based Spatial Clustering of Applications with Noise (HDBSCAN) algorithm \citep{McInnes2017}. The HDBSCAN algorithm, which is an extension of Density-Based Spatial Clustering of Applications with Noise, is a density-based clustering algorithm designed to identify density-connected regions by calculating distances between data points. 
The HDBSCAN algorithm introduces a hierarchical structure by constructing a distance matrix and generating a directed weighted graph, which it traverses to assign data points to clusters.
HDBSCAN exhibits a more powerful clustering performance and is capable of discovering clusters of arbitrary shape and size. 
The clustering results are shown in Figure \ref{fig:figure3},
where the FRB samples are divided into two clusters. 
One of the two clusters, which contains a significant proportion of repeaters, 
is labeled as the repeater cluster (more than $10\%$ of the FRBs are repeaters), while another cluster is termed as the non-repeater cluster. 
Although the rest of the FRBs in the repeater cluster have only 
one burst detected, their features are close to the repeating FRBs in the 
selected parameter space. We label such FRBs as the repeater candidates. 
Table \ref{tab:tab3} lists the detailed content of the two clusters. 
On the other hand, there is one repeating FRB categorized in the 
non-repeater cluster. We further discuss the verification of results in Section \ref{ssec:Verif}.

We further explore potential correlations between the embedding space clustering results and FRB morphology as well as host galaxy properties. Based on \citet{2021ApJ...923....1P}, we randomly select several FRBs from the first CHIME/FRB catalog and categorize them as simple broadband bursts, simple narrowband bursts, complex broadband bursts with multiple peaks, complex narrowband bursts with multiple peaks, and downward-drifting bursts for comparison with Figure 3. The results indicate that broadband bursts consistently belong to the non-repeater cluster, while narrowband bursts are always classified within the repeater cluster. However, there is no clear correlation between the two clusters and complex, simple, or downward-drifting bursts. Host galaxy properties are also crucial for understanding the origins and characteristics of FRBs. According to the recent statistical results by \citet{2024arXiv241109203M}, nine FRBs in the first CHIME/FRB catalog are localized, including five repeaters and four non-repeaters. We compare the host galaxy morphological properties and stellar masses of these nine FRBs with the t-SNE dimensionality reduction results. The comparison indicates that there is no significant difference in host galaxy properties between repeaters and non-repeaters, consistent with the conclusions of \citet{2024arXiv241109203M}.

In the following sections, we refer to the FRB catalog sample categorized
in the t-SNE embedding space as the grouped sample (\textbf{S\textsubscript{grp}}), which 
consists of 94 repeaters, 163 repeater candidates, and 337 non-repeaters.

\begin{table}\scriptsize
\begin{center}
\caption{A summary of the FRBs numbers in \textbf{S\textsubscript{grp}}.}
\label{tab:tab3}
\tabletypesize{\scriptsize}
\begin{tabular}{lcc}
\hline
\bm{} & Repeater Cluster  &    Non-repeater Cluster  \\
\hline
Repeater   & 93     & 1  \\
Apparent non-repeater  & 163   & 337  \\          
Total    & 256     & 338 \\
\hline
\end{tabular}
\end{center}
\end{table}

In the initial \textbf{S\textsubscript{org}} catalog, there are $94$ repeaters and the repeater fraction is
about $ 94/ (94+506) \times 100\% = 15.8\%$. 
Considering the 163 repeater candidates that t-SNE predicted, the repeater fraction 
is increased to $(94+163)/(94+506) \times 100\% = 42.8\%$. The result is at the same level of 
the theoretical predictions \citet{2022MNRAS.509.1227C,2023MNRAS.522.4342Y}.
This suggests that the true fraction of repeaters should be much greater than 
that reported by current observations. 
Consequently, this means that the severe contamination in \textbf{S\textsubscript{org}} hinders research into the physical mechanisms of repeaters and non-repeaters.

\subsection{Feature Importance} \label{ssec:feature}

\begin{figure*}
\centering
\begin{minipage}{0.24\textwidth}
    \centering
    \includegraphics[angle=0,scale=0.178]{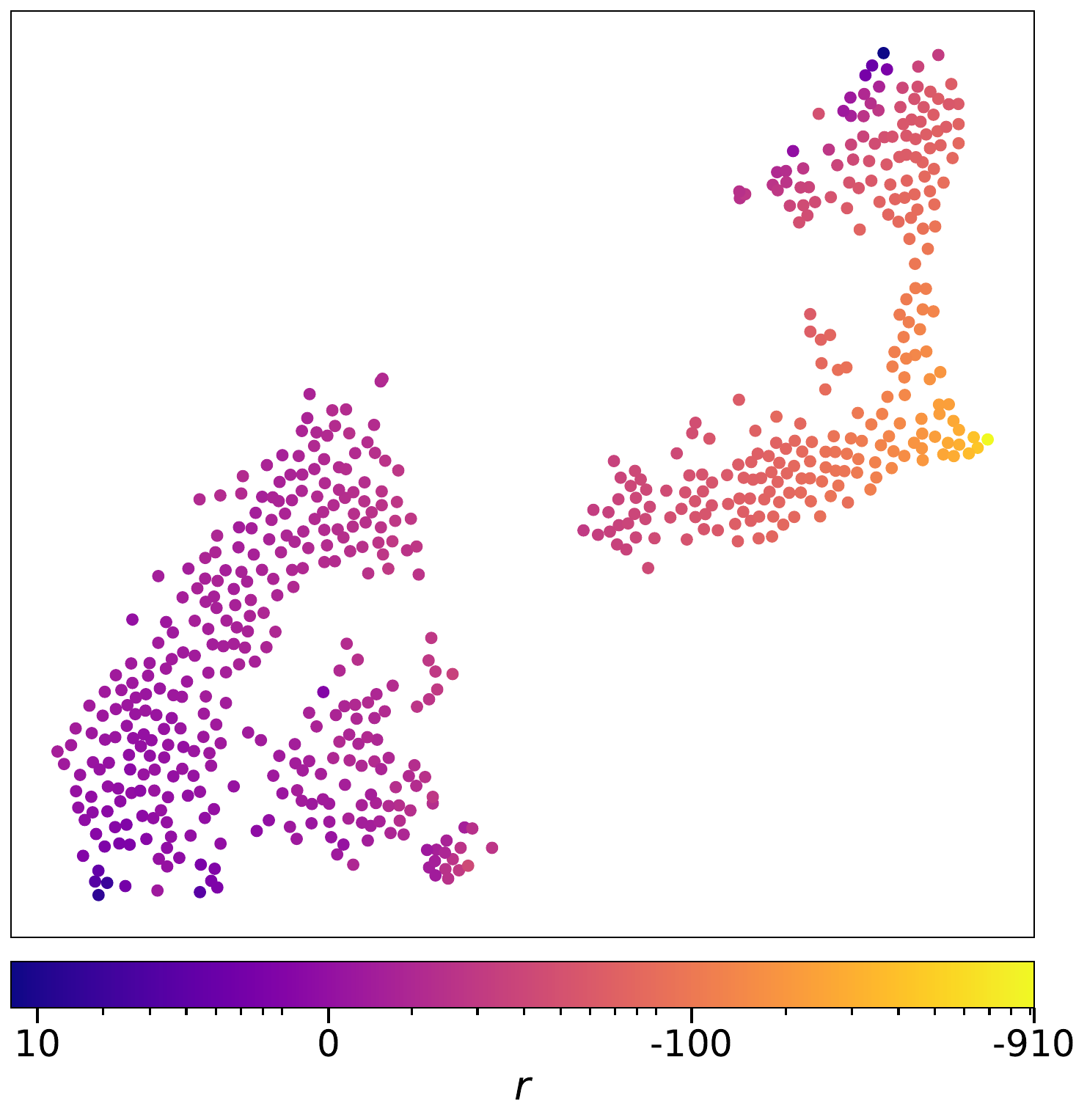}
\end{minipage}%
\begin{minipage}{0.24\textwidth}
    \centering
    \includegraphics[angle=0,scale=0.178]{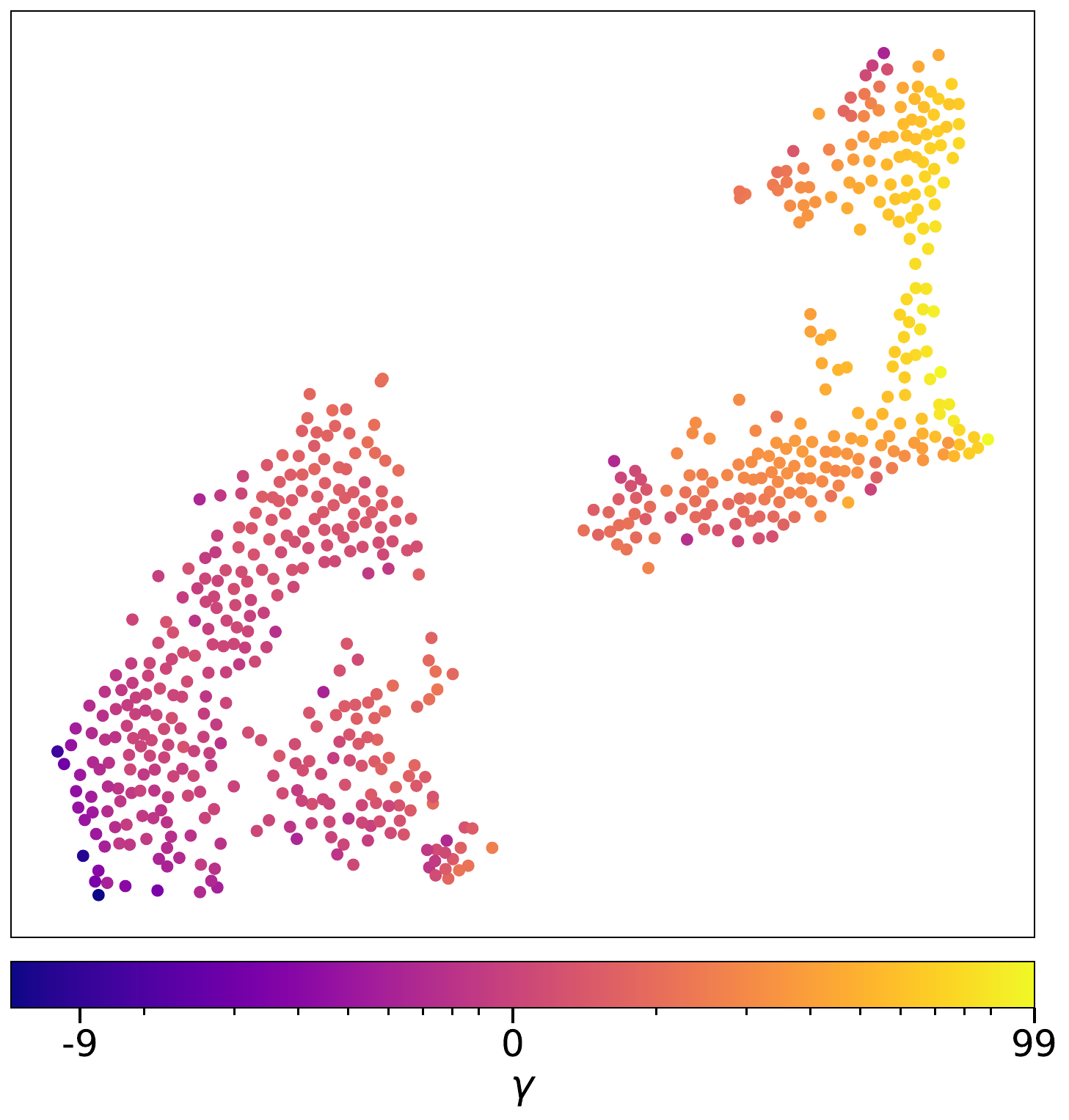}
\end{minipage}%
\begin{minipage}{0.24\textwidth}
    \centering
    \includegraphics[angle=0,scale=0.178]{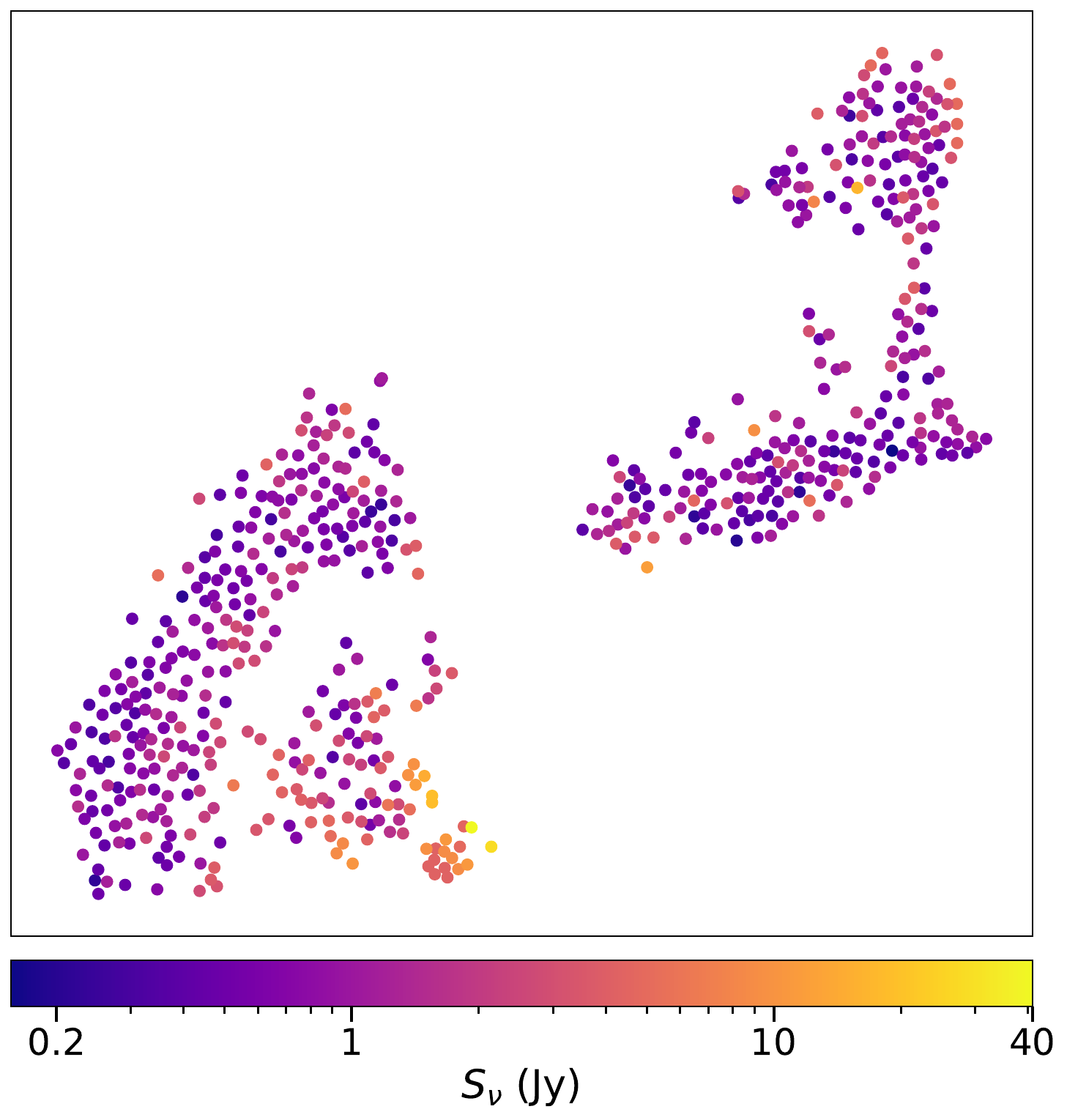}
\end{minipage}%
\begin{minipage}{0.24\textwidth}
    \centering
    \includegraphics[angle=0,scale=0.178]{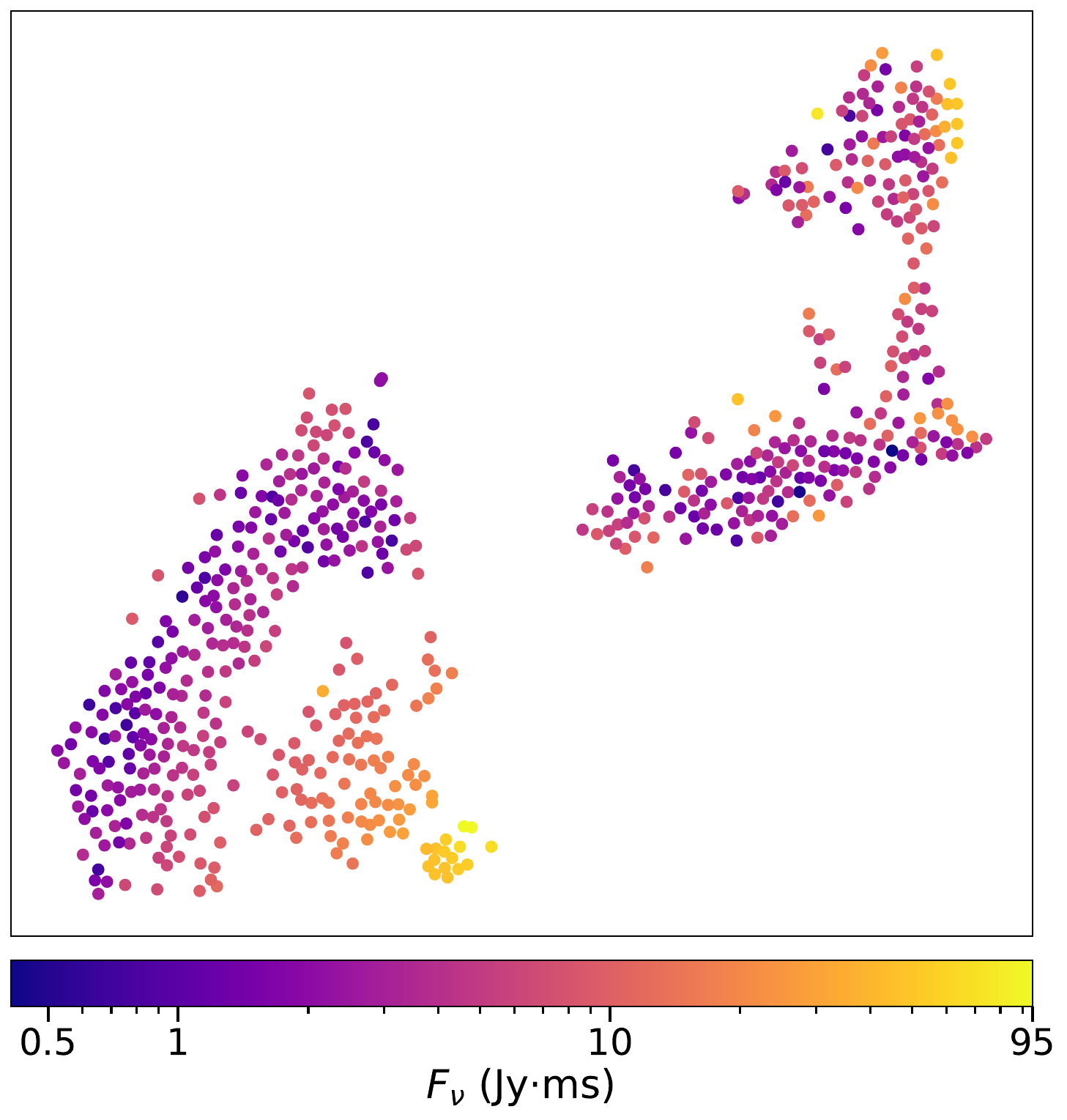}
\end{minipage}
\vskip\baselineskip
\begin{minipage}{0.26\textwidth}
    \centering
    \includegraphics[angle=0,scale=0.178]{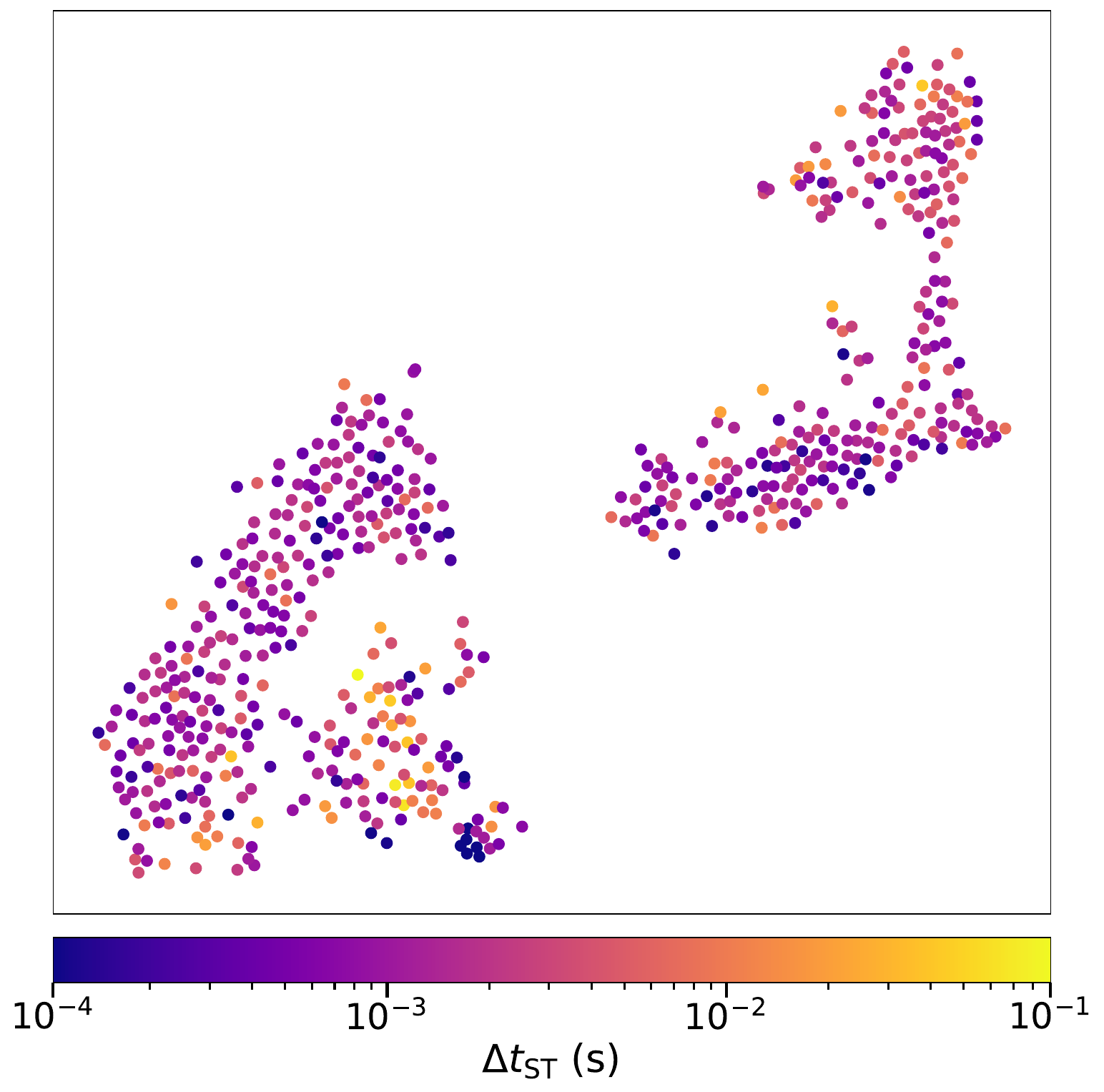}
\end{minipage}%
\begin{minipage}{0.26\textwidth}
    \centering
    \includegraphics[angle=0,scale=0.178]{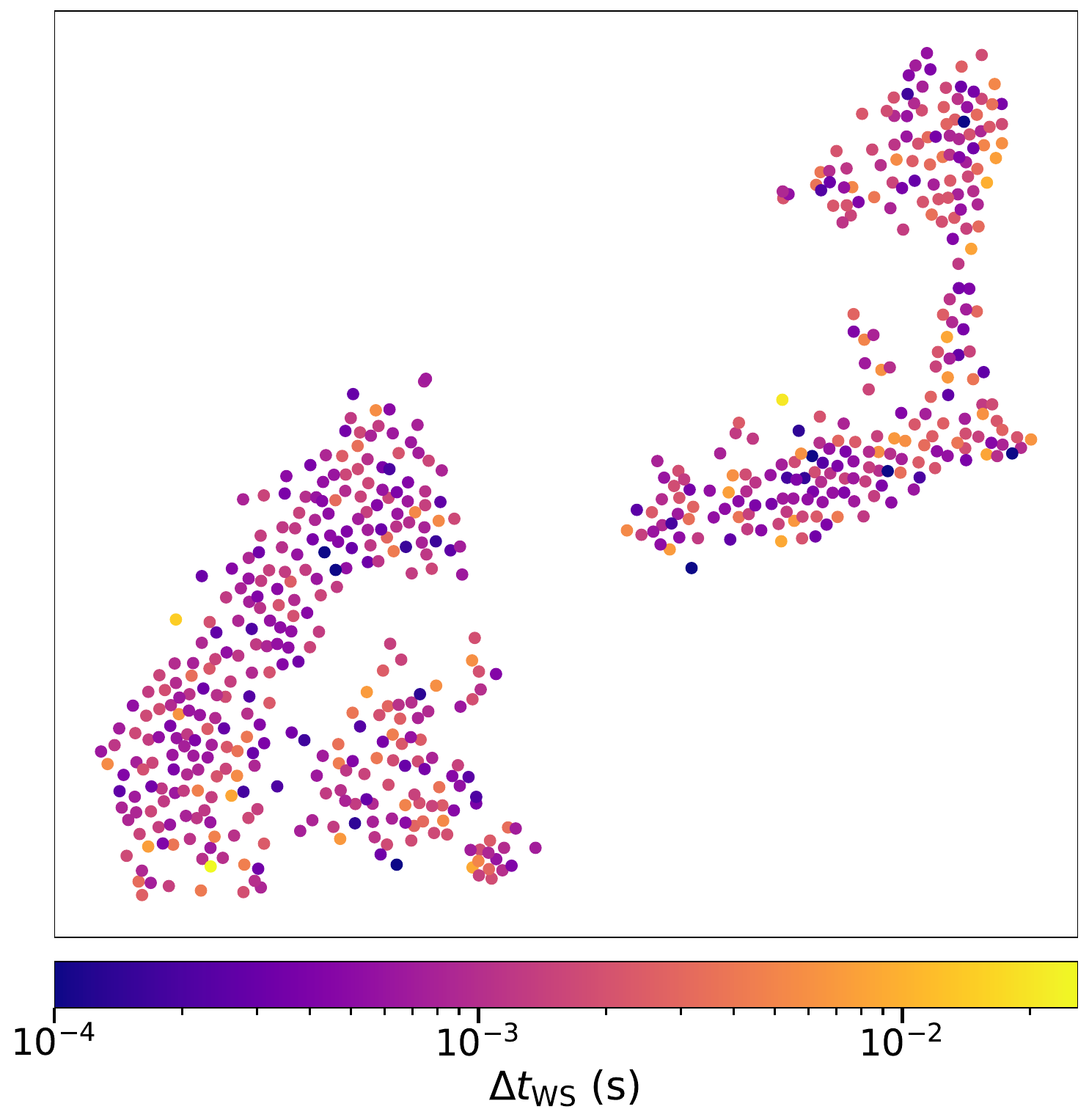}
\end{minipage}%
\begin{minipage}{0.26\textwidth}
    \centering
    \includegraphics[angle=0,scale=0.178]{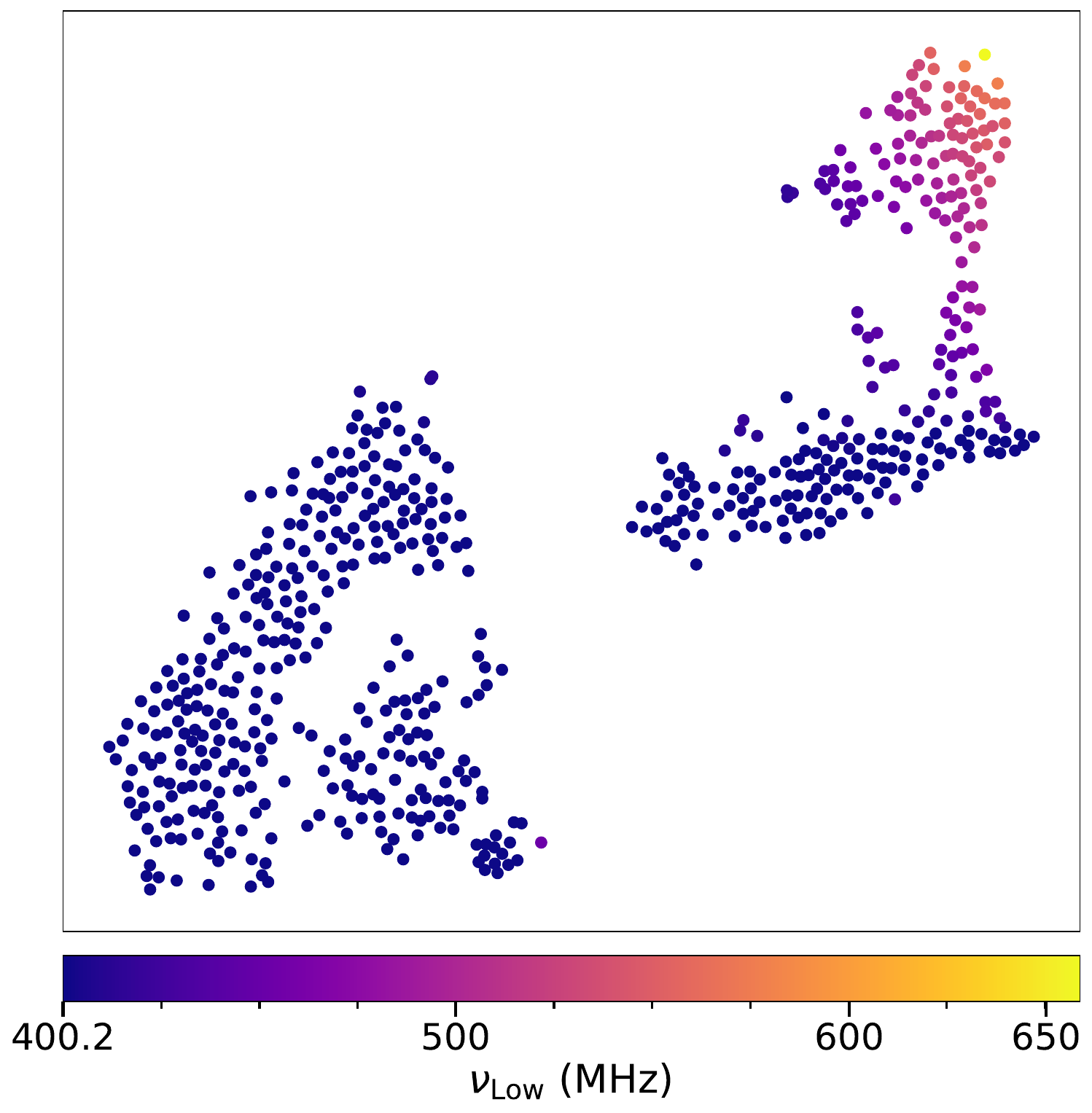}
\end{minipage}
\caption{t-SNE embedding space, colored based on $r$, $\gamma$, $S_{\nu}$, $F_{\nu}$, $\Delta t_\mathrm{ST}$, $\Delta t_\mathrm{WS}$, and $\nu_\mathrm{Low}$.}
\label{fig:figure4}
\end{figure*}

In the previous section, we show the results of the unsupervised FRB classification 
in the t-SNE embedding space based on seven observation parameters. 
To gain a deeper understanding of the FRB classification process,
we explore which observable parameters of FRBs are more significant in 
machine learning classification. 
For better visualization, we apply color coding for each FRB of the two clusters 
based on the corresponding values of $r$, $\gamma$, $S_{\nu}$, $F_{\nu}$, $\Delta t_\mathrm{ST}$, $\Delta t_\mathrm{WS}$, and $\nu_\mathrm{Low}$, respectively. As shown in Figure \ref{fig:figure4}, the values of $r$, $\gamma$, and $\nu_\mathrm{Low}$ show 
obvious gradients in the t-SNE embedding space. Other physical parameters exhibit no noticeable regular variation between these two clusters.
By visual inspection, t-SNE has roughly approximated the FRBs into two clusters according to $r$ and $\gamma$. However, there is no clear boundary for $r$ and $\gamma$ alone between the two clusters. 
Specifically, the $r$ values of some FRBs in the repeater cluster are greater than 
some non-repeater clusters'.

To further quantitatively investigate the impact of observational parameters on machine learning classification, we define performance metrics that represent scores as:
\begin{equation}
{\rm Recall} = \frac{T_{\rm P}}{T_{\rm P} + F_{\rm N}}, 
\end{equation}
where $T_{\rm P}$, i.e. true positive, denotes the number of repeaters that fall in the 
repeater clusters, and $F_{\rm N}$, i.e. false negative, denotes the number of repeaters
that fall in the non-repeater clusters. 
We adopt a model validation technique named permutation feature importance \citep{2010Bioinformatics...26..10,2023MNRAS.522.4342Y}. 
Specifically, we sequentially shuffle the values of the seven FRB observed parameters and
apply the value-shuffled data to the same t-SNE classification process. 
By shuffling the values of one particular FRB observation parameter, 
the similarity information carried by such parameters is disrupted, leading to a loss in recall.
The permutation feature importance is quantified by the recall loss rate caused:
\begin{equation}
{\rm Loss_{rate}} = \frac{\rm Recall_{original} - \rm Recall_{shuffled}}{\rm Recall_{original}} \times 100 \%, 
\end{equation}
where Recall$_{\rm original}$ denotes the initial recall value before any parameter is shuffled, serving as the baseline for performance evaluation, and Recall$_{\rm shuffled}$ denotes the recall value after a specific parameter has been shuffled and used to the same t-SNE model. 
A parameter with greater Loss$_{\rm rate}$ indicates an important feature for identifying repeaters. 
Finally, we repeat the procedure 10 times for each parameter and calculate the average Loss$_{\rm rate}$. 
Therefore, we combine HDBSCAN clustering with the definition of recall as a performance metric to ultimately assess the extent to which the t-SNE model relies on specific features.

\begin{figure}
\centering
\includegraphics[angle=0,scale=0.5]{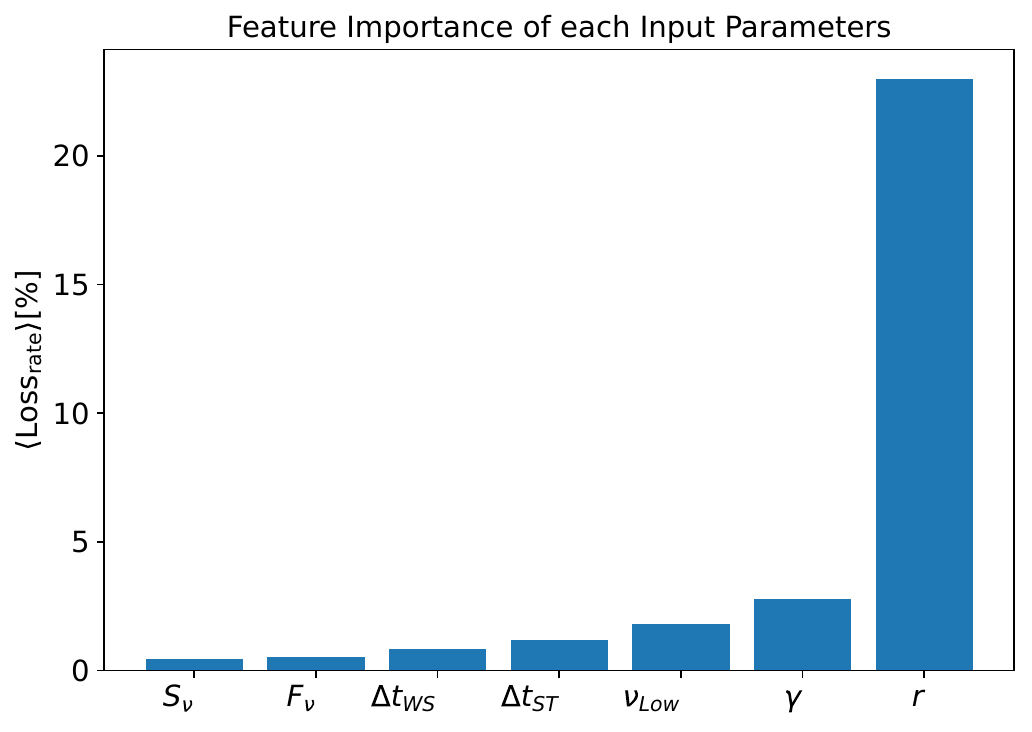}
\caption{The result of permutation feature importance for our t-SNE model.
\label{fig:figure5}}
\end{figure}

Figure \ref{fig:figure5} illustrates the importance of the permutation feature, i.e. the recall loss rate of the seven FRB observational parameters. The results show that the parameters of $r$ and $\gamma$ are 
the most important, i.e. the loss caused by the permutation of $r$ reaches $22.98\%$ and 
the second-ranking $\gamma$ causes a $2.77\%$ loss. 
The importance of the other observational parameters shows relatively minor differences.
Therefore, we show that in the unsupervised classification of FRBs, 
the parameter $r$ plays a pivotal role in differentiating repeaters and non-repeaters.

\begin{figure*}[htbp]
\centering
\includegraphics[angle=0, scale=0.20]{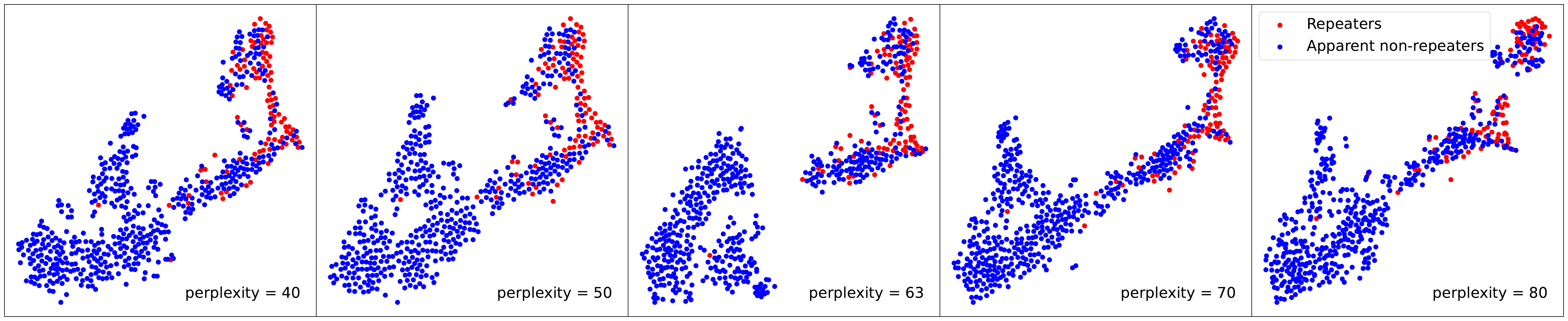}
\\
\includegraphics[angle=0, scale=0.20]{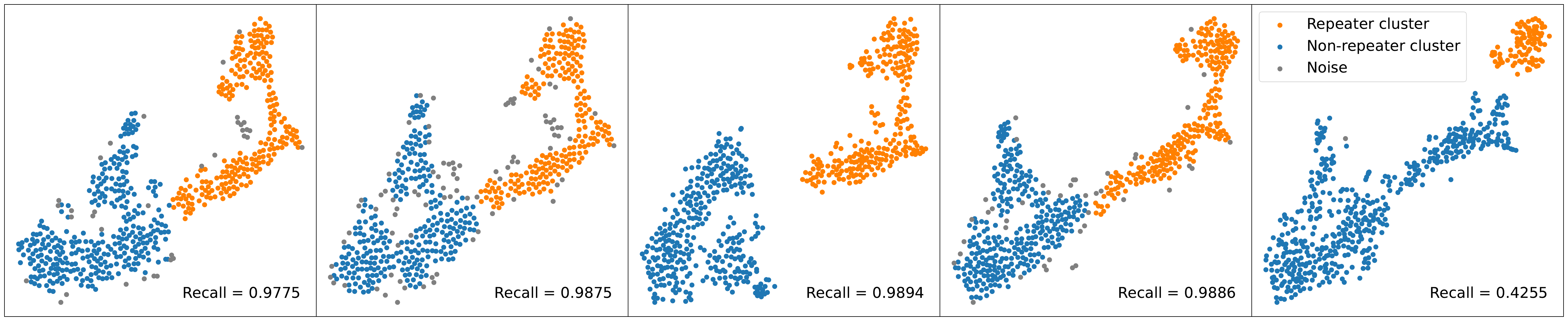}
\caption{t-SNE embedding results (upper panel) and HDBSCAN clustering results (lower panel) for different perplexity values.}
\label{fig:figure6}
\end{figure*}

\subsection{Dimensionality Reduction Stability} \label{ssec:Stability}

t-SNE presents a challenge in finding low-dimensional representations of data, as different initializations often lead to varying embedding results. To obtain more meaningful representations of the data structure, a common approach is to run t-SNE multiple times, which allows for the selection of an optimal range of parameters and, in turn, improves the overall stability of the embeddings. In t-SNE embeddings, the perplexity parameter determines the cluster sizes, making the selection of an appropriate perplexity value crucial for t-SNE results stability. To evaluate the stability of the embeddings and classification outcomes, we conduct experiments with different perplexity values and analyze their effects. As shown in the upper panel of Figure \ref{fig:figure6}, the separation between repeaters and the majority of apparent non-repeaters in the low-dimensional embeddings remains highly consistent across varying perplexity values, with only minor variations in the local structures. 

To further quantify the impact of different perplexity values on performance, we combine the embeddings with HDBSCAN clustering analysis and used the recall metric from Equation (1) as the evaluation criterion. In the lower panel of Figure \ref{fig:figure6}, the recall values derived under consistent HDBSCAN clustering parameters (\textit{min\_cluster\_size} = 100, and \textit{min\_samples} = 5) remain above 0.9 when the perplexity value is within the range of 40–70, indicating a high level of classification stability. However, as the perplexity value increases to 80, the embedding structure becomes less effective at accurately distinguishing between repeaters and non-repeaters, leading to a drop in recall below 0.5. 

\begin{figure*}[ht!]
\centering
\begin{minipage}[t]{0.292\linewidth} 
    \centering
    \includegraphics[width=\linewidth, trim=0 0 0 0, clip]{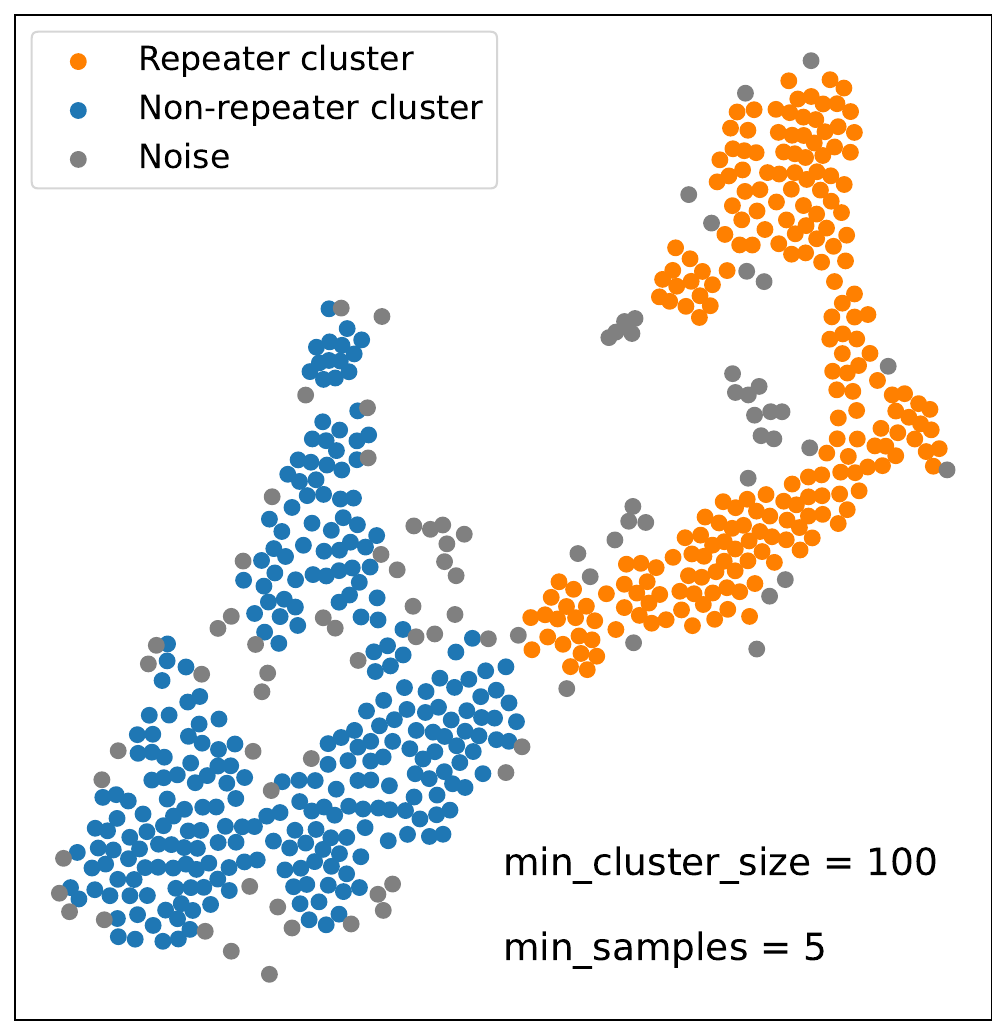}
\end{minipage}\hspace{0pt}%
\begin{minipage}[t]{0.30\linewidth}
    \centering
    \includegraphics[width=\linewidth, trim=0 0 0 0, clip]{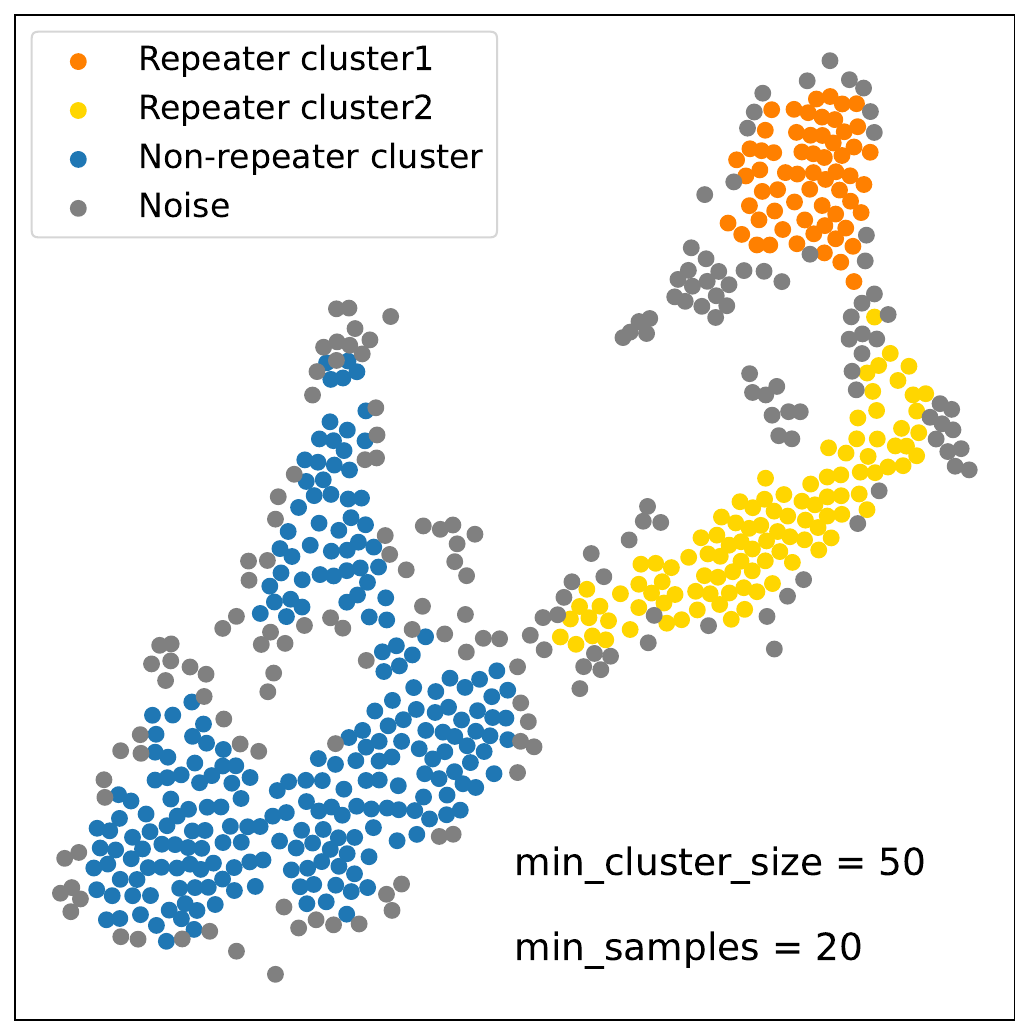}
\end{minipage}\hspace{0pt}%
\begin{minipage}[t]{0.3\linewidth}
    \centering
    \includegraphics[width=\linewidth, trim=0 0 0 0, clip]{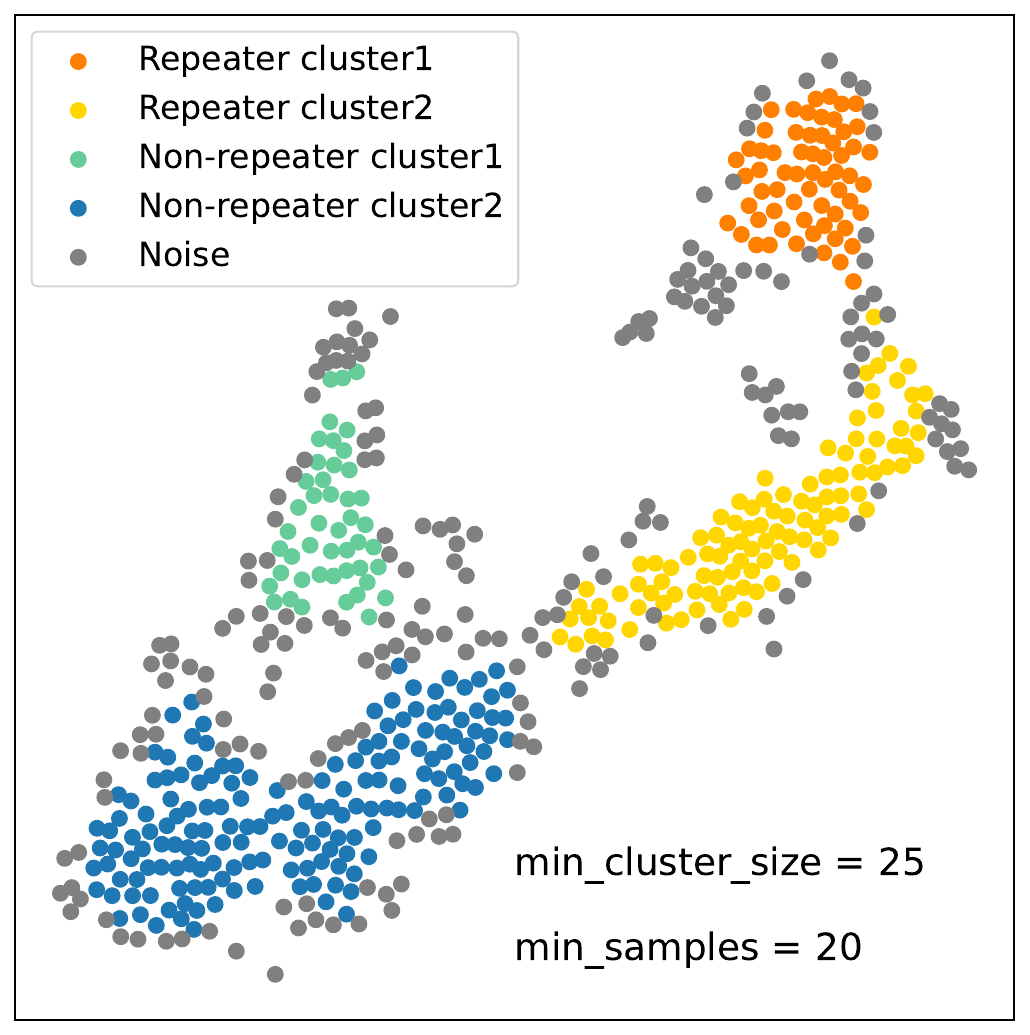}
\end{minipage}
\caption{Clustering results of the t-SNE embedding space at perplexity = 50 for different HDBSCAN parameters.
}
\label{fig:figure7}
\end{figure*}

On the other hand, varying the HDBSCAN clustering parameters also affects the final results. 
We observe that there might be substructures within the repeater cluster and non-repeater cluster, e.g., the repeater cluster shows a distinct trend of separation along the vertical direction. As shown in Figure \ref{fig:figure4}, there is a noticeable difference in the values of $r$ and $\nu_\mathrm{Low}$ between the upper and lower parts of the repeater cluster. Similarly, the non-repeater cluster exhibits some separation along the horizontal direction, with the primary difference between the two subgroups being observed in the values of $F_{\nu}$.
In the topological structures obtained with different perplexity values, i.e. perplexity = 50 as shown in Figure \ref{fig:figure7}, subcategories can be identified. 
With different parameter settings of HDBSCAN, both the repeater and non-repeater clusters exhibited subcluster results. 
The emergence of these subclusters suggests that there may be multiple emission mechanisms or different evolutionary stages within repeating and non-repeating bursts \citep{2024PASA...41...90L}. This observation warrants further investigation, and we plan to explore this issue in detail with larger sample sizes in future studies.

In the embedding space with perplexity = 63, a wide range of adjustments to HDBSCAN's main parameters (\textit{min\_cluster\_size} ranged from 2 to 100, and \textit{min\_samples} ranged from 5 to 50) consistently resulted in the identification of only two clusters. This highlights the significance and stability of the two-cluster result, while also indicating that the density levels of the substructures are insufficient to form reliable clusters. Moreover, the number of indeterminate noise points varies across embedding spaces with different perplexity values. Consequently, we fine-tuned the perplexity value within the range of 50--70, finding that the most meaningful low-dimensional structure, with the highest recall of 0.9894, is achieved when perplexity = 63.

\subsection{Results Verification} \label{ssec:Verif}

\begin{figure}
\centering
\includegraphics[angle=0,scale=0.28]{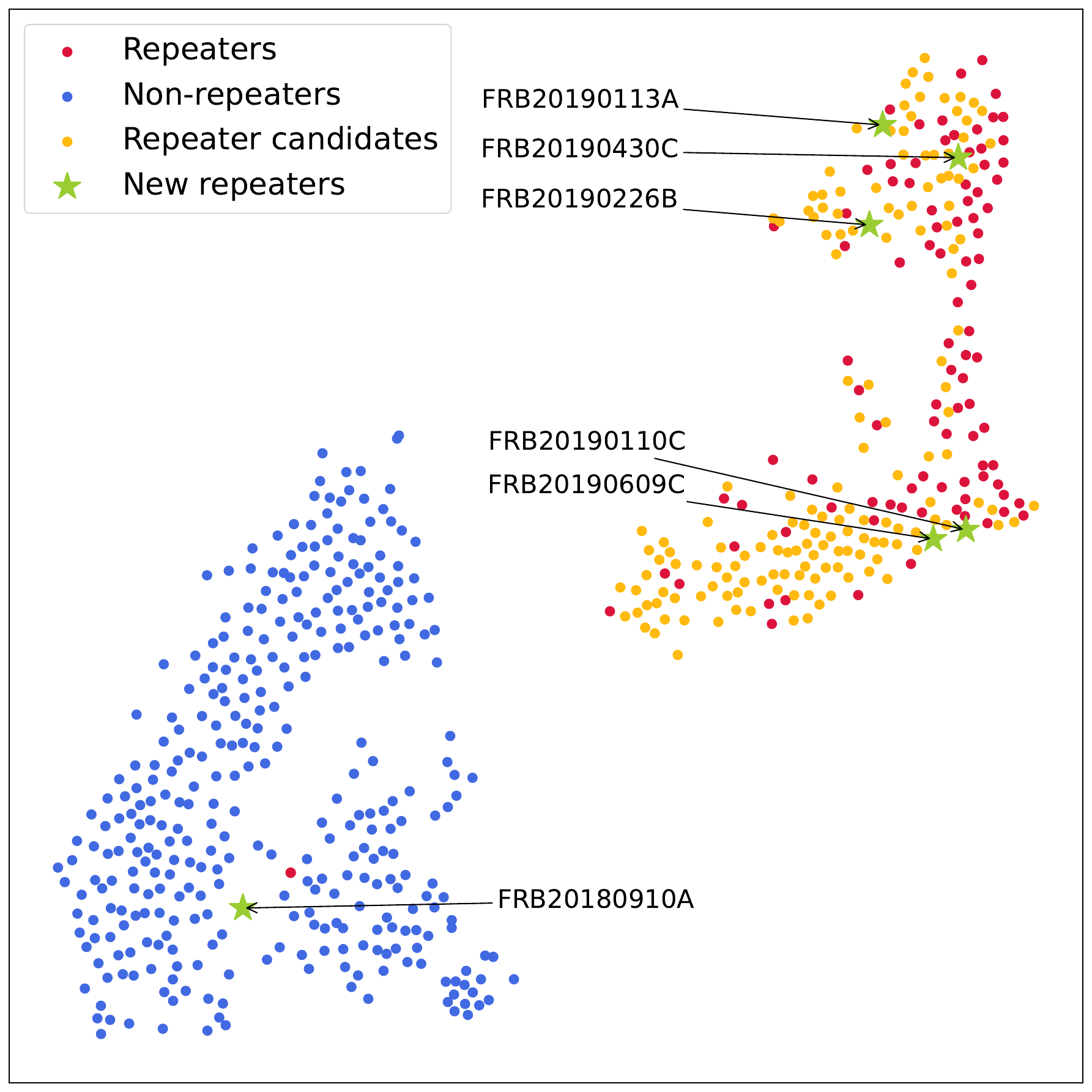}
\caption{The FRB \textbf{S\textsubscript{grp}} catalog distribution in the t-SNE embedding space.
The repeaters are marked in red, the repeater candidates are marked in orange,
and the non-repeaters are marked in blue.
The six newly confirmed repeaters in the work of \citet{2023ApJ...947...83C}
are marked with the green stars.
\label{fig:figure8}}
\end{figure}

When investigating dimensionality reduction and clustering of FRBs, we pay special 
attention to individual repeaters with unusual behavior.
These specific repeaters show similarities with the non-repeater cluster regarding the observational parameters, prompting further investigation into their unique properties. 
In this section, we analyze the detailed observational properties of the unique repeaters, 
i.e. FRB20181030A and FRB20180910A, to examine their similarities with general non-repeaters
and explore the potential reasons for this anomalous behavior.

As shown in Figure \ref{fig:figure2}, there is one repeating FRB (FRB20181030A, represented with a red dot) falling within the non-repeater cluster.
We compare the observational information and the dynamic spectrum images of FRB20181030A with its subsequent repeat, FRB20181030B, neither of which exhibits any sub-bursts. 
In Table \ref{tab:tab4}, we provide a comparison of basic observational parameters, including sky localization, DM values, and other fundamental observables. While these bursts do not initially appear to originate from two distinct sources, given the localization uncertainties, we cannot entirely rule out this possibility. 
By inspecting their dynamic spectrum images, both bursts intuitively appear to be repeating bursts
with narrowband emission. The narrowband feature of FRB20181030B (see \url{https://www.chime-frb.ca/catalog/FRB20181030B}) is particularly prominent. 
However, we note that FRB20181030A (see \url{https://www.chime-frb.ca/catalog/FRB20181030A}) has
a low SNR and its frequency spectrum is noise-dominated. 
This may have affected the reliability of its derived parameters, resulting in FRB20181030A being incorrectly fitted as broadband emission. Therefore, we suspect that the classification of the repeater FRB20181030A as a non-repeater may be due to data quality issues rather than its intrinsic properties.

\begin{table*}\scriptsize
\begin{center}
\caption{Comparison of observational parameters for FRB20181030A and FRB20181030B.}
\label{tab:tab4}
\setlength{\tabcolsep}{2pt}
\begin{tabular}{lccccccccc}
\hline
TNS Name & Repeater Name & R.A. (J2000) & Decl. (J2000) & DM & $S_{\nu}$ & $F_{\nu}$ & $\gamma$ & $r$ & SNR \\
 &  & (°) & (°) & (pc cm$^{-3}$) & (Jy) & (Jy ms) &  &  &  \\
\hline
FRB20181030A & FRB20181030A & $158.35\pm0.180$ & $73.79\pm0.310$ & 103.396 & 4.30 & 8.20 & $3.50\pm3.40$ & $-3.10\pm3.70$ & 10.80 \\
FRB20181030B & FRB20181030A & $158.35\pm0.180$ & $73.79\pm0.310$ & 103.643 & 2.30 & 6.90 & $22.90\pm2.50$ & $-43.20\pm4.00$ & 32.90 \\
\hline
\end{tabular}
\end{center}
\end{table*}

To verify the reliability of our repeater candidates' identification, we validate our 
predictions using an updated repeater catalog \citep{2023ApJ...947...83C}. 
There are $25$ repeating FRBs reported and six of them were previously observed as non-repeaters
in the first CHIME/FRB Catalog. 
These six newly confirmed repeaters are marked with green stars in Figure \ref{fig:figure8}.

As shown in Figure \ref{fig:figure8}, five of them are located in the repeater cluster
on the right side, which were successfully identified beforehand as repeater candidates
using our method. It demonstrates reliable accuracy in identifying potential repeater candidates. 
More importantly, this also indicates that the seven observational parameters we selected in Section \ref{ssec:data desc} are indeed key features for machine learning in distinguishing between repeaters and non-repeaters.
 
There is one newly confirmed repeater, i.e. FRB20180910A, located in the non-repeater 
cluster in the t-SNE embedding space, which is also marked with the green star. 
By analyzing the dynamic spectrum of FRB20180910A, we discover that this repeater exhibits a broadband emission characteristic similar to non-repeaters, which is unique compared to the narrowband emission typically observed in repeaters. To investigate that individual bursts, such as FRB20180910A, do not significantly bias our results, we repeat the t-SNE embedding and clustering analysis after excluding this burst. The analysis confirmed that the removal of FRB20180910A did not alter the general clustering structure and the main conclusions of this study. This consistency demonstrates the robustness of our approach to outlier removal.

The physical explanation of such an unusual broadband emission repeater is not yet clear. 
Whether this is a manifestation of different states in repeater activity or 
caused by observational effects during propagation requires further investigation
with more observational evidence in the future.

\subsection{Parameters Distribution} \label{ssec:paramt}

\begin{figure*}
    \centering
    \includegraphics[angle=0, scale=0.30]{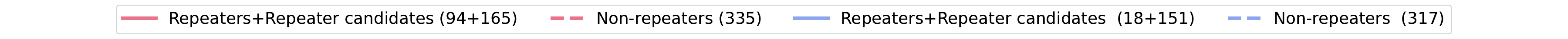}
    
    \begin{minipage}{0.23\textwidth}
        \centering
        \includegraphics[angle=0, scale=0.20]{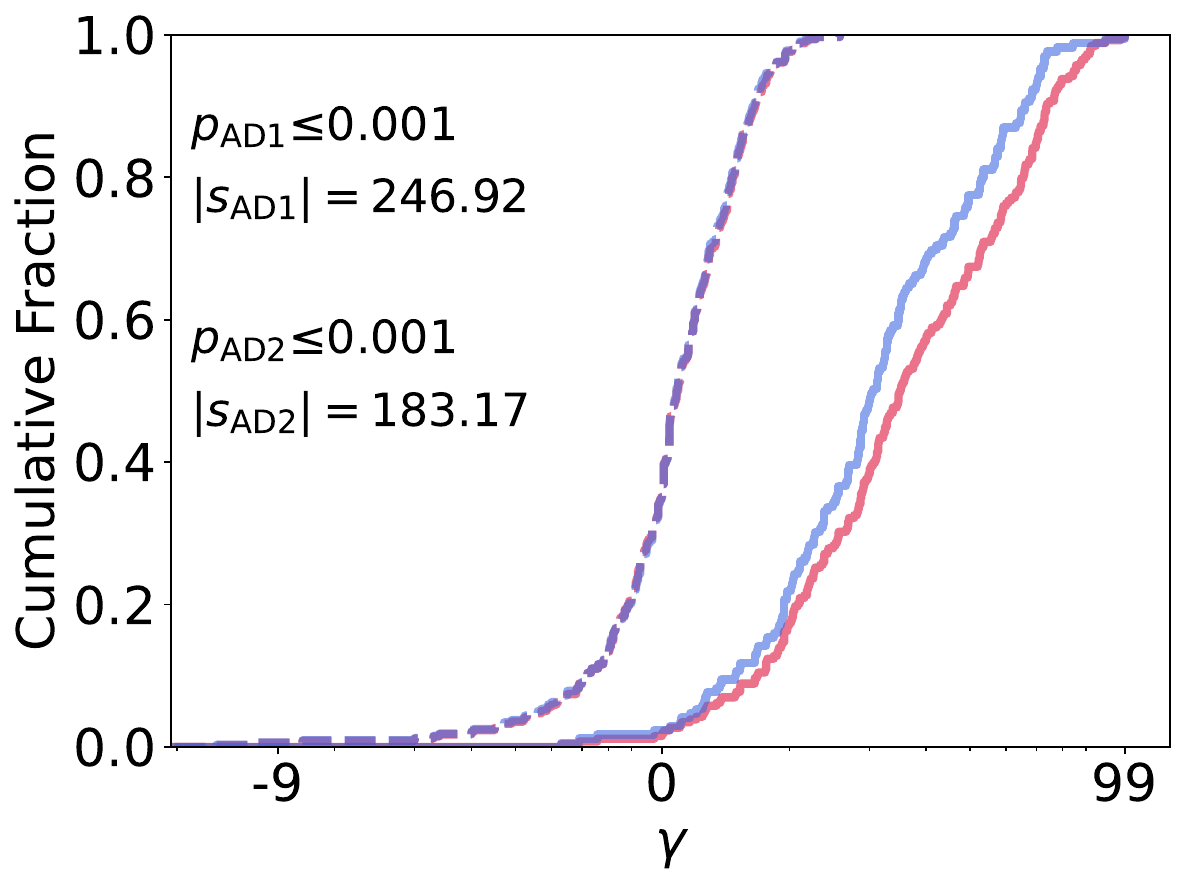}
    \end{minipage}%
    \begin{minipage}{0.23\textwidth}
        \centering
        \includegraphics[angle=0, scale=0.20]{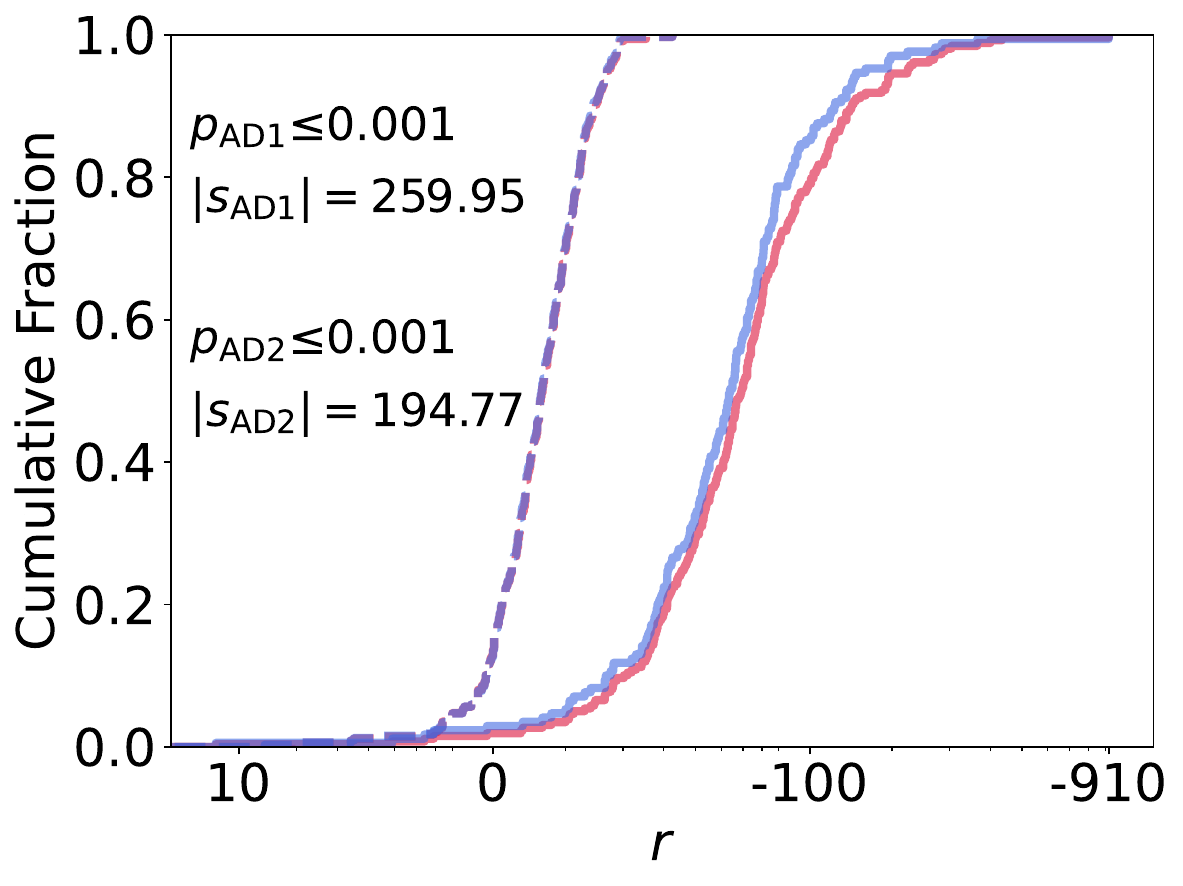}
    \end{minipage}%
    \begin{minipage}{0.23\textwidth}
        \centering
        \includegraphics[angle=0, scale=0.20]{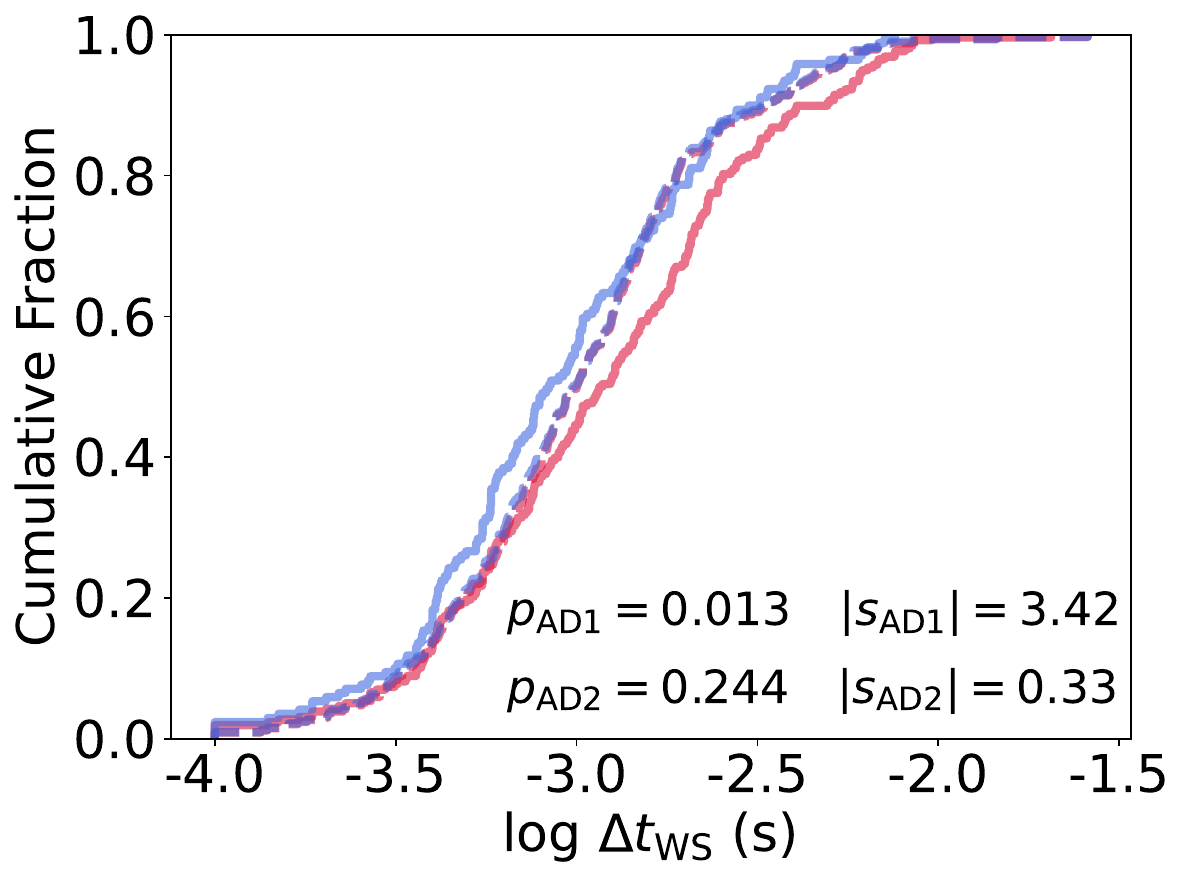}
    \end{minipage}%
    \begin{minipage}{0.23\textwidth}
        \centering
        \includegraphics[angle=0, scale=0.20]{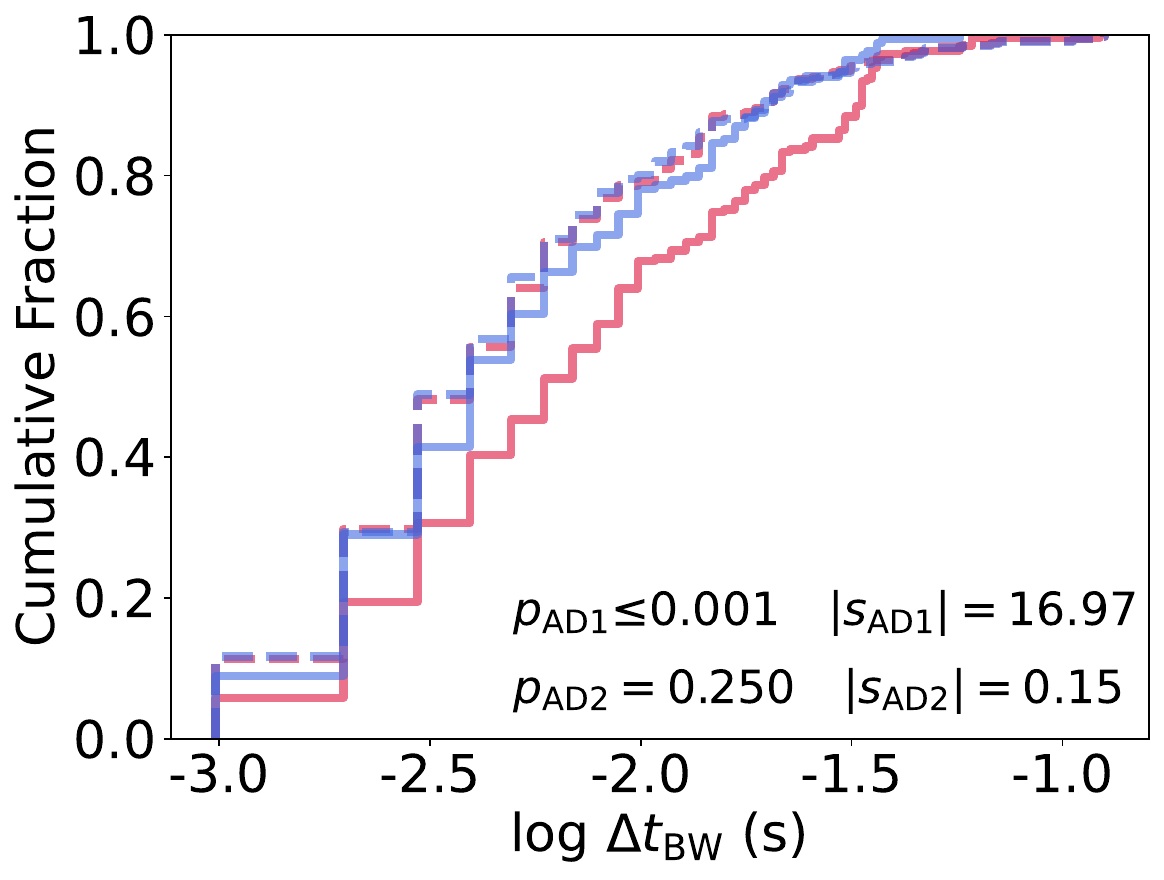}
    \end{minipage}
    
    \begin{minipage}{0.23\textwidth}
        \centering
        \includegraphics[angle=0, scale=0.20]{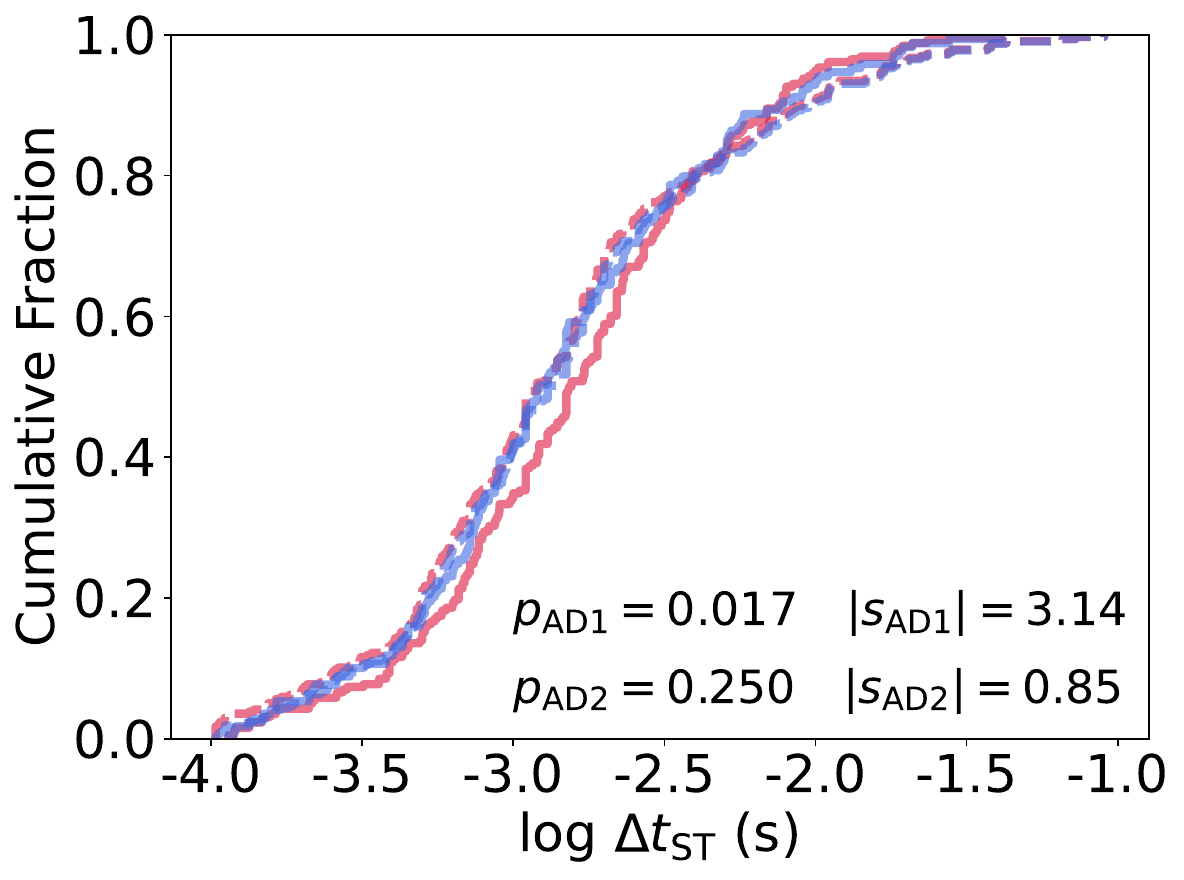}
    \end{minipage}%
    \begin{minipage}{0.23\textwidth}
        \centering
        \includegraphics[angle=0, scale=0.20]{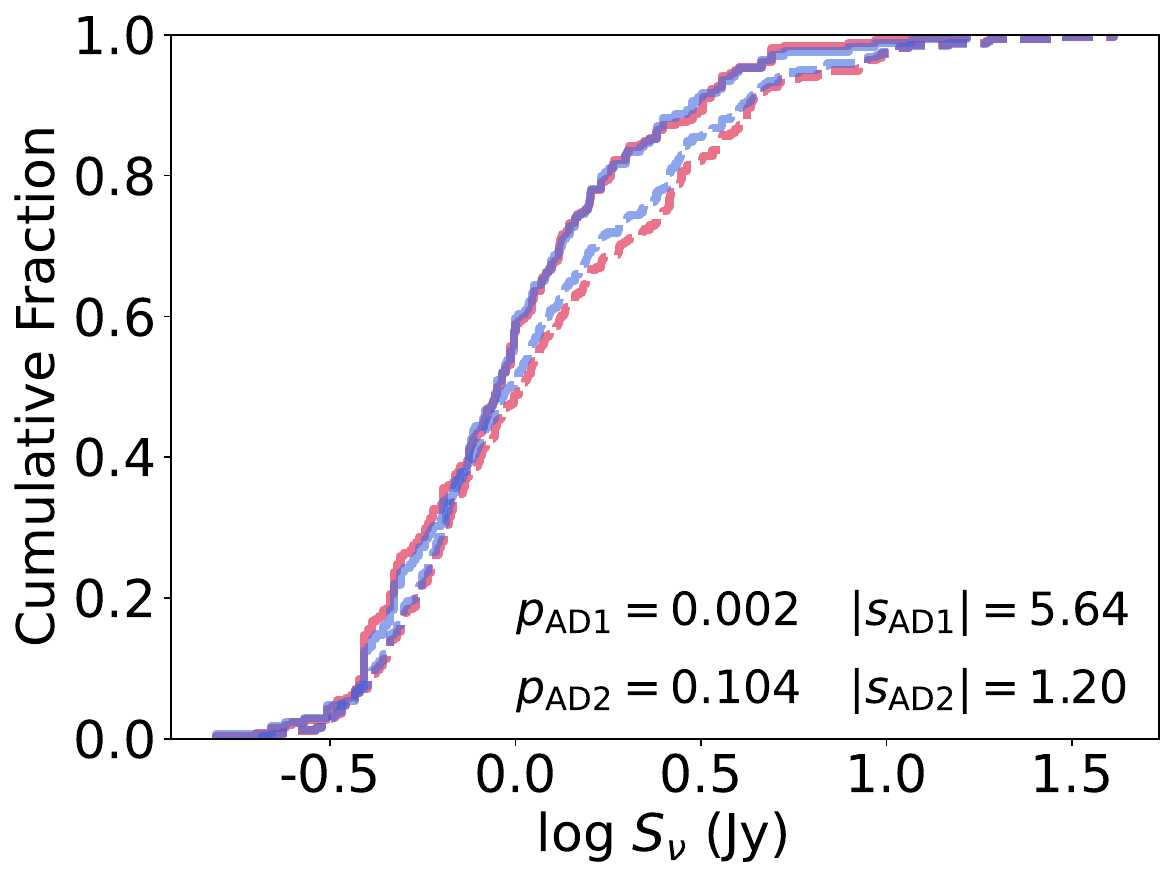}
    \end{minipage}%
    \begin{minipage}{0.23\textwidth}
        \centering
        \includegraphics[angle=0, scale=0.20]{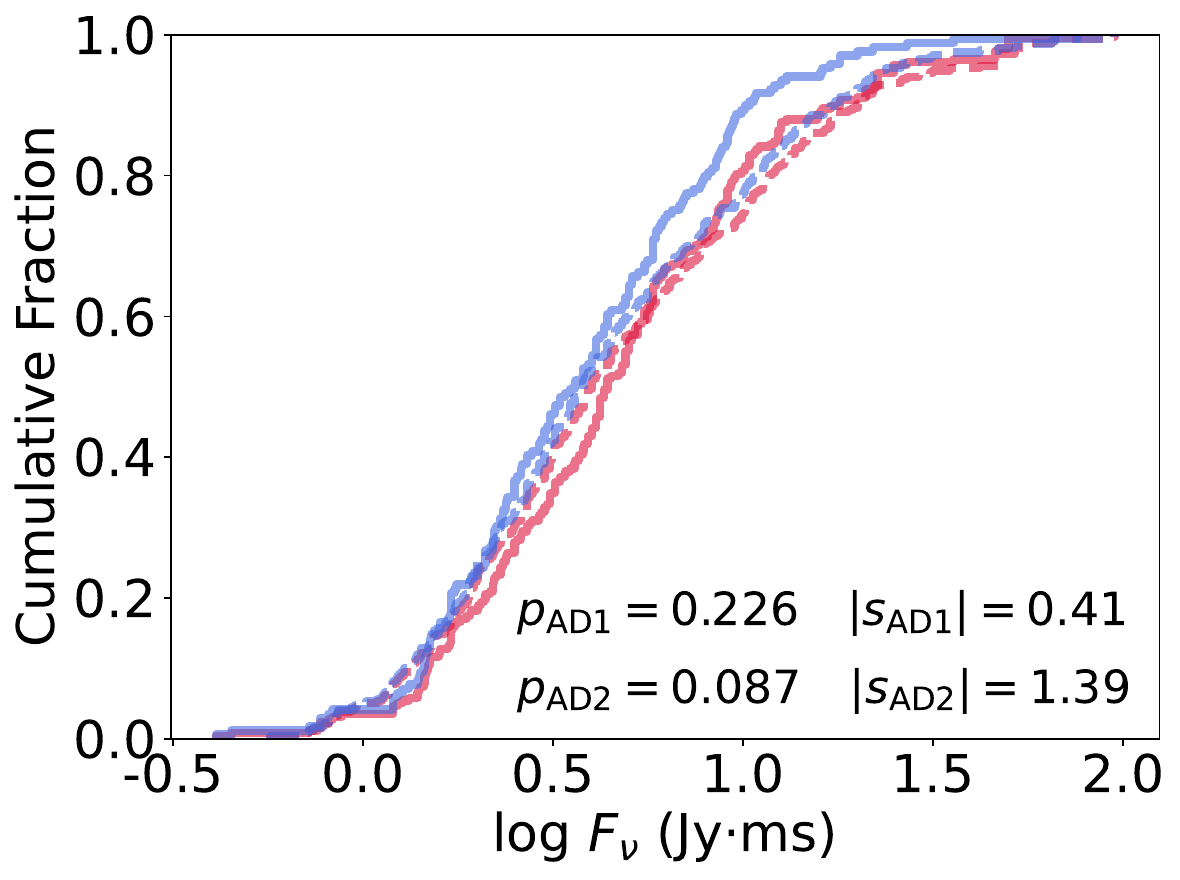}
    \end{minipage}%
    \begin{minipage}{0.23\textwidth}
        \centering
        \includegraphics[angle=0, scale=0.20]{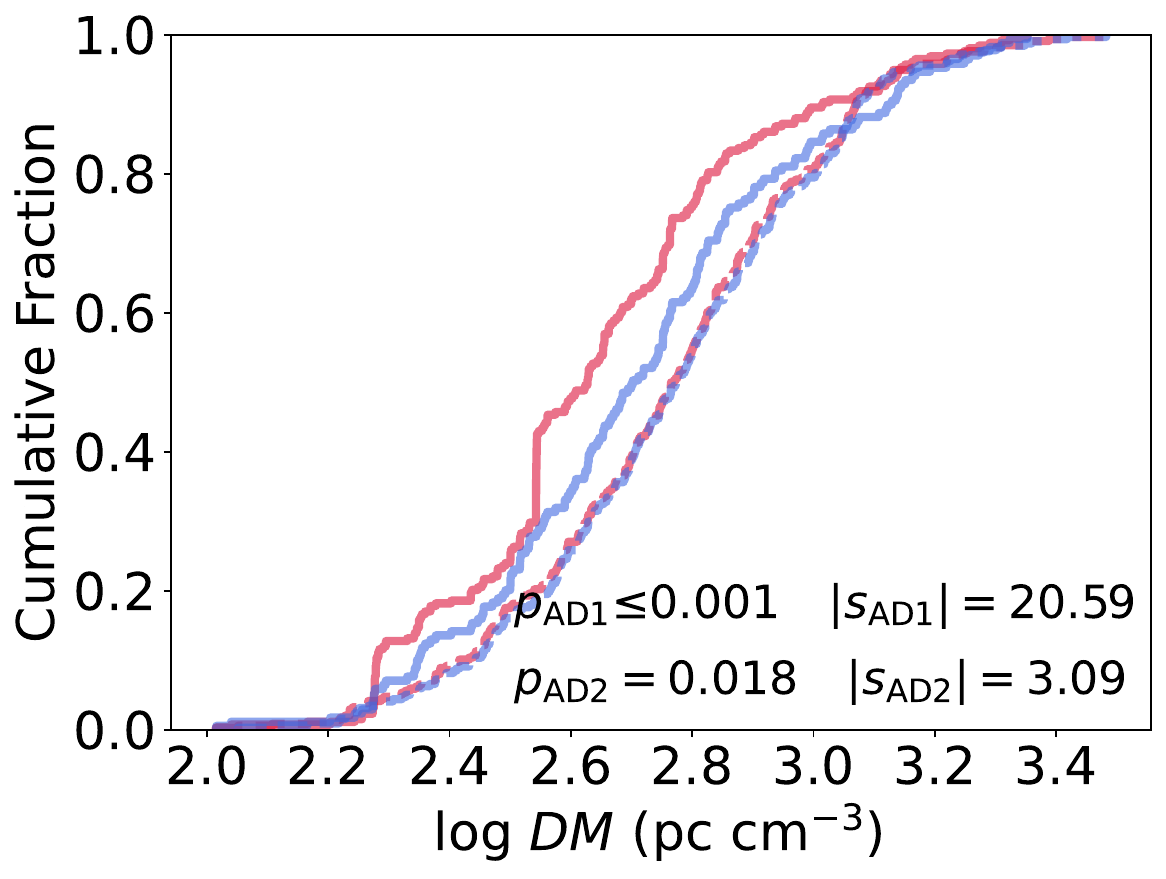}
    \end{minipage}
    
    \begin{minipage}{0.23\textwidth}
        \centering
        \includegraphics[angle=0, scale=0.20]{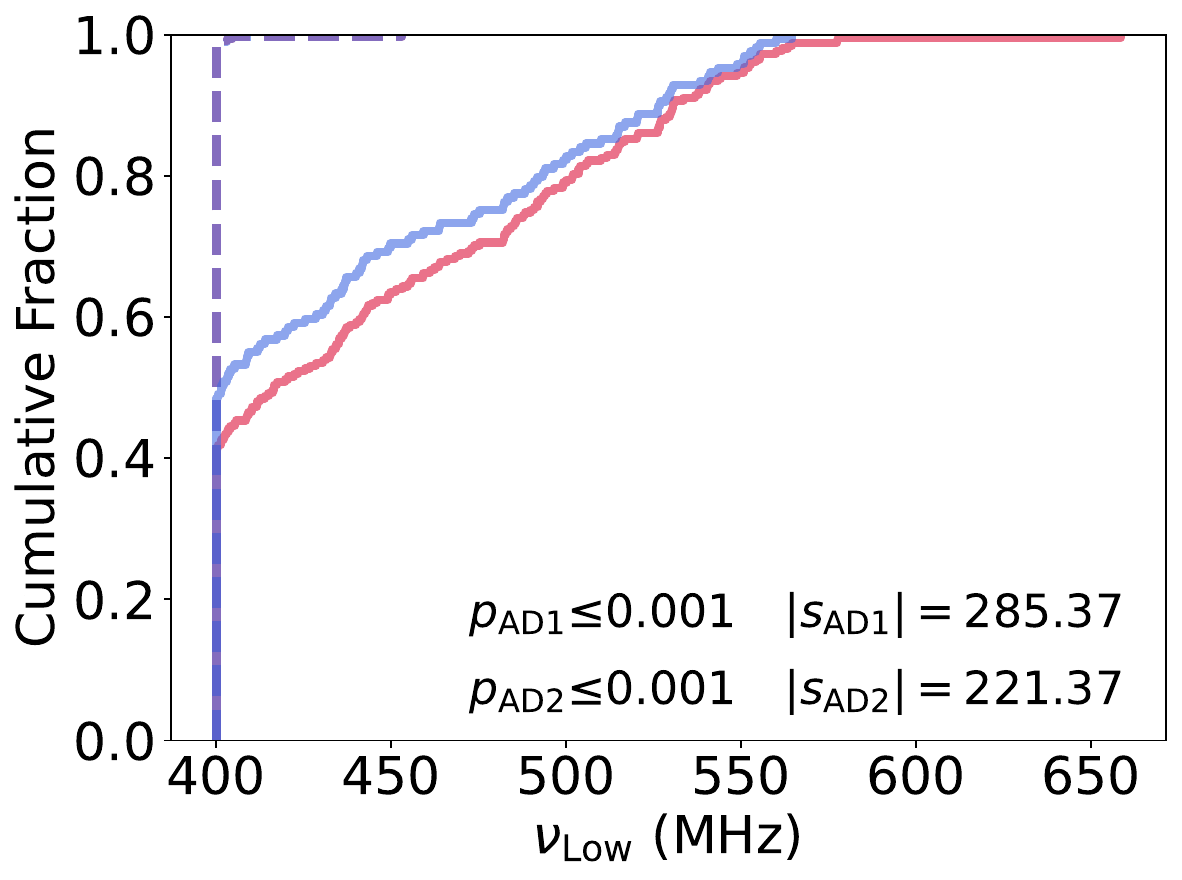}
    \end{minipage}%
    \begin{minipage}{0.23\textwidth}
        \centering
        \includegraphics[angle=0, scale=0.20]{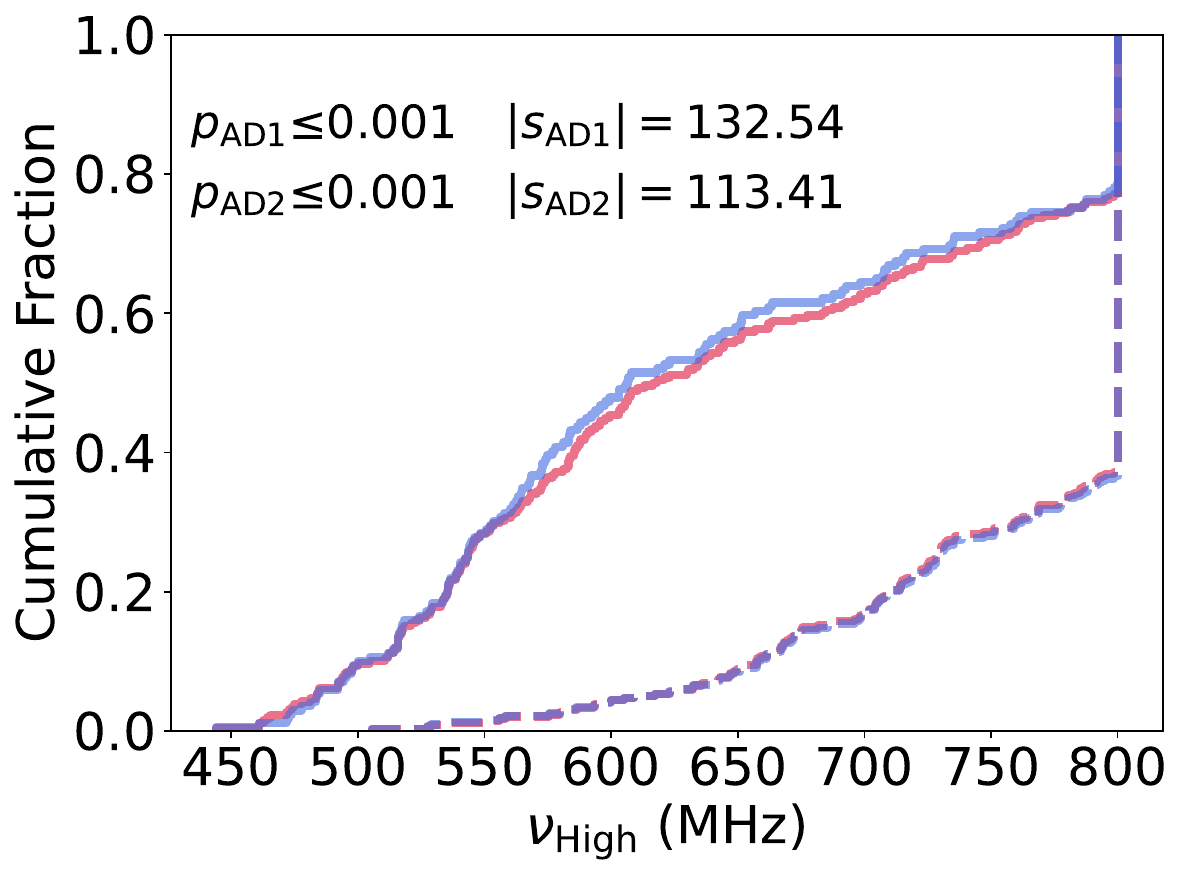}
    \end{minipage}%
    \begin{minipage}{0.23\textwidth}
        \centering
        \includegraphics[angle=0, scale=0.20]{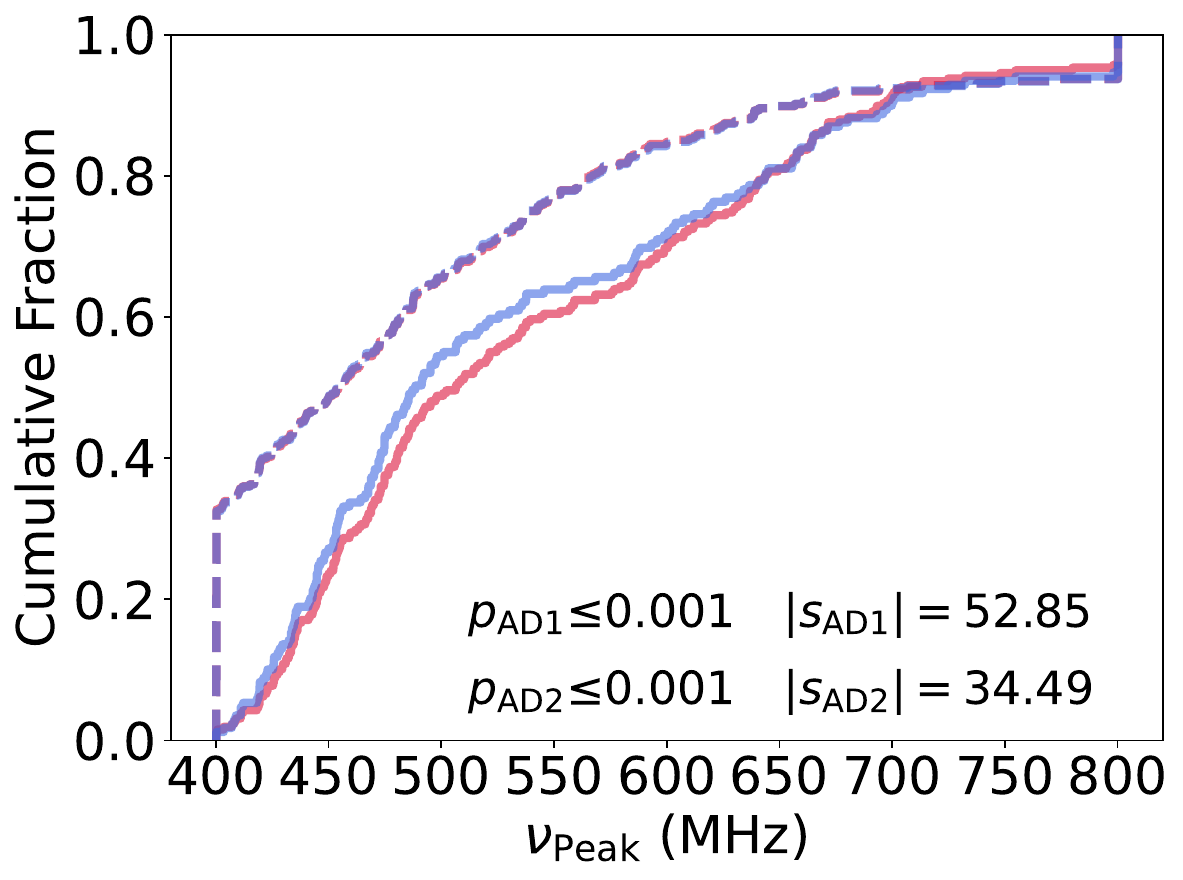}
    \end{minipage}%
    \begin{minipage}{0.23\textwidth}
        \centering
        \includegraphics[angle=0, scale=0.20]{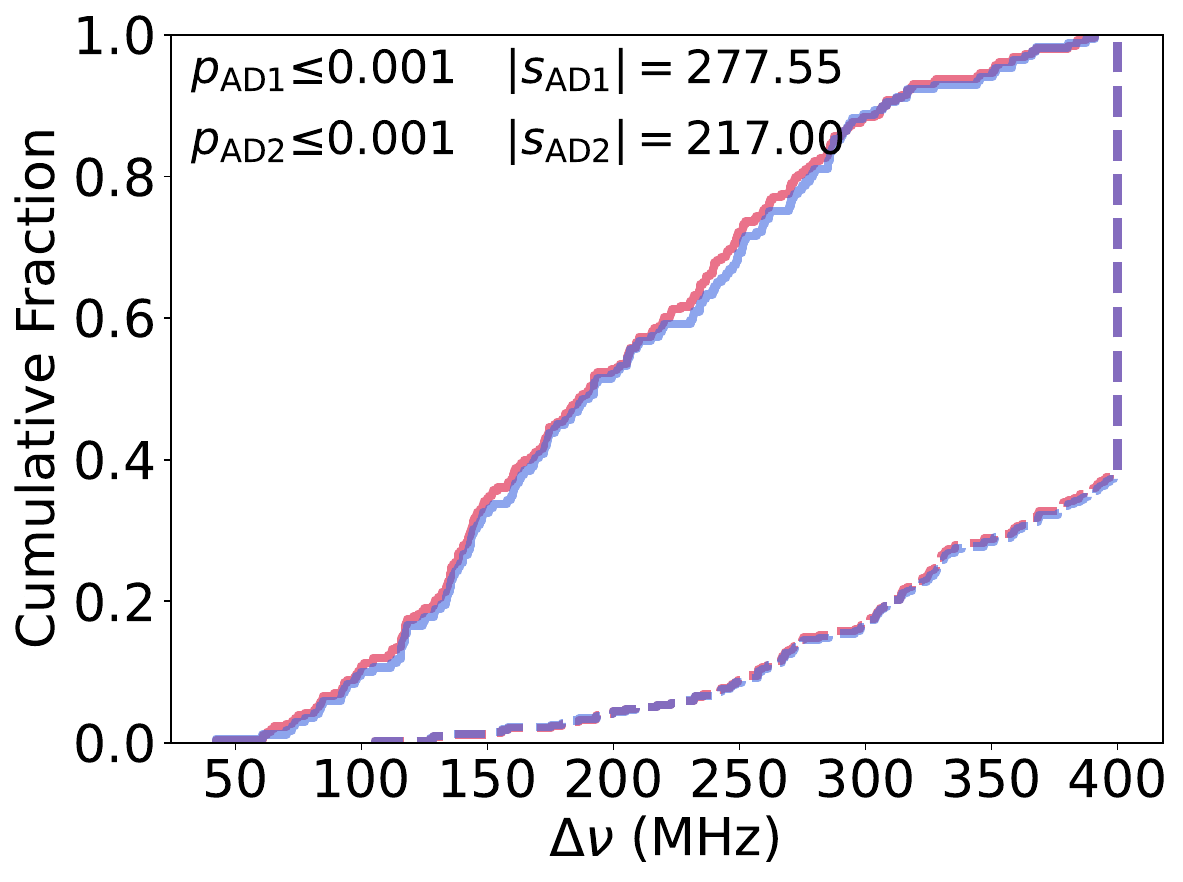}
    \end{minipage}

    \caption{The cumulative distributions of observational parameters of the first CHIME/FRB catalog. $p_\mathrm{AD1}$ represents the approximate p-value of the AD test conducted on the \textbf{S\textsubscript{grp}} sample, while $p_\mathrm{AD2}$ corresponds to the results after excluding all sub-bursts and multiple repeaters from the same repeating source. These p-values indicate the statistical significance of the differences between the two sample distributions, while the normalized k-sample AD test statistic ($s_\mathrm{AD}$) quantifies the extent of these differences. Note that p-values are floored at 0.001.}
    \label{fig:figure9}
\end{figure*}

To intuitively investigate the feature differences among repeaters, repeater candidates, and non-repeaters, we perform the empirical cumulative distribution function (ECDF) analysis and the Anderson--Darling (AD) test \citep{anderson1954test} on the FRB feature parameters, combining repeaters and repeater candidates into one group and plotting them together, while plotting non-repeaters separately. In the following study, we refer to both the repeaters and the repeater candidates as the repeaters for analysis. The null hypothesis of the AD test is that repeaters and non-repeaters are drawn from the same population. If the p-value of the AD test is smaller than 0.05, we reject the null hypothesis and conclude that repeaters and non-repeaters are statistically different.

The ECDFs and AD test results for the reclassified S\textsubscript{grp} sample (comprising 94 repeaters, 163 repeater candidates, and 337 non-repeaters) are displayed in Figure \ref{fig:figure9}, with the solid and dashed red lines representing repeaters and non-repeaters, respectively. The AD test results are indicated by $p_\mathrm{AD1}$, which represents the statistical significance of the difference between the two sample distributions, and $s_\mathrm{AD1}$, the normalized test statistic quantifying the extent of this difference. Furthermore, to better investigate the differences in parameter distributions between repeating and non-repeating FRB sources, we conduct an additional analysis after excluding all sub-bursts from the sample. Specifically, for bursts with multiple sub-bursts, we retain only the event with \textit{sub\_num} = 0. Additionally, we compare only the first-detected repeater events for each repeating source since these events were triggered at the nominal threshold, thus minimizing the statistical complications arising from multiple events per source. The resulting filtered sample includes 18 repeaters, 151 repeater candidates, and 317 non-repeaters. The ECDF analysis and AD tests for the sub-burst-excluded sample are also presented in Figure \ref{fig:figure9}, with the solid and dashed blue lines representing repeaters and non-repeaters, respectively, and $p_\mathrm{AD2}$ and $s_\mathrm{AD2}$ showing the test results.

We illustrate the $11$ physical parameters of FRB features introduced in 
Section \ref{ssec:data desc}. 
Additionally, we also plot the distribution of frequency bandwidth, i.e. 
$\Delta \nu=\nu_\mathrm{High}-\nu_\mathrm{Low}$. 
Except for $\nu_\mathrm{Low}$, $\nu_\mathrm{High}$, $\nu_\mathrm{Peak}$, and $\Delta \nu$, the ECDFs are uniformly binned on a logarithmic scale. Unless otherwise specified, the following analysis of parameter distribution differences defaults to referencing $p_\mathrm{AD2}$ and $s_\mathrm{AD2}$.

As shown in Figure \ref{fig:figure9} 
the cumulative distributions of $\gamma$ and $r$ show significant differences 
between repeaters and non-repeaters. Moreover, the AD test results for these two parameters show $p_\mathrm{AD2} \leq 0.001$, with $s_\mathrm{AD2}$ values of 183.17 and 194.77, respectively, indicating that $\gamma$ and $r$ have  intrinsically different distributions between repeaters and non-repeaters.
The spectral shape for a single burst is characterized by
\citep{2021ApJS..257...59C}:
\begin{equation}
  F\left(\gamma, r\right)=\left(\frac{f}{f_{0}}\right)^{\gamma + r \ln \left( \frac{f}{f_{0}}\right)},
\label{equ:equ1}
\end{equation}
where $F$ represents the spectral shapes, $\gamma$ is the spectral index, $r$ is the spectral running, and
$f_{0} = 400.2$ MHz is the CHIME observation lower frequency limit, respectively. 
The spectral running $r$ characterizes the turnover of the frequency spectrum and is thus 
related to the burst frequency bandwidth, i.e. $r \sim 0$ indicates a broadband emission. 
As shown in Figure \ref{fig:figure9}, the non-repeaters exhibit $r \sim 0$, 
whereas repeaters tend to exhibit $r$ < 0. Such differences indicate that the 
non-repeaters have broadband emission, while repeaters are more likely to be narrowband. 
This is consistent with the studies in the literature
\citep{2020ApJ...891L...6F,2021ApJ...923....1P,2021ApJS..257...59C,2023ApJ...947...83C}.

Similarly, the cumulative distribution of bandwidth $\Delta \nu$ further confirms this distinction, with repeaters favoring narrowband emission and non-repeaters inclined toward broadband emission ($p_\mathrm{AD2} \leq 0.001$, $s_\mathrm{AD2} = 217$).

Previous studies consistently indicate that repeaters generally exhibit wider pulses than non-repeaters, both in terms of observed and intrinsic widths \citep{2016ApJ...833..177S,2019ApJ...885L..24C,2020ApJ...891L...6F,2021ApJS..257...59C,2022ApJ...926..206Z,2023ApJ...947...83C}. However, we find no significant difference 
between the two populations, with $p_\mathrm{AD2}=0.244$. It may indicate that the difference in burst width 
currently exhibited by the two populations is due to the sample mixing issue. 
The potential repeaters may have been misclassified, causing statistical discrepancies 
that were not considered in the previous study. 
Furthermore, \citet{2020MNRAS.497.3076C} explained this difference by the selection effect of 
beam emission. This needs to be gradually confirmed with the publication of a larger 
sample of observations in the future.

Notably, we observe that for DM, $p_\mathrm{AD2} = 0.018$, indicating a difference between repeaters and non-repeaters. Although previous studies have not observed this difference \citep{2019ApJ...885L..24C,2020ApJ...891L...6F,2021ApJ...923....1P,2021ApJS..257...59C}, but \citet{2023ApJ...947...83C} reported $p_\mathrm{AD} = 0.0112$ after analyzing more observed repeaters, finding a significant difference in DM between these two populations. The nature of this difference remains unclear, as it is susceptible to observational selection effects \citep{2021A&A...647A..30G}. Future studies will require more detailed population modeling to investigate this further \citep{2023ApJ...947...83C}.
 
In addition, Figure \ref{fig:figure9} shows that the AD test results for $S_{\nu}$, $\Delta t_\mathrm{WS}$, $\Delta t_\mathrm{BW}$, and $\Delta t_\mathrm{ST}$ reveal different distribution differences depending on whether sub-bursts and multiple repeaters are considered. This may be influenced to varying degrees by the instrument's time resolution and sample size limitations, which are areas worthy of further investigation in the future.

To further study the intrinsic properties of FRBs, it is essential to delve into the distribution of 
their isotropic energy ($E_{\rm{iso}}$), isotropic peak luminosity ($L_{\rm{iso}}$), 
and brightness temperature ($T_{\rm{B}}$). Since most FRBs are unlocalized and lack precise redshift $z$ measurements, we first estimate $z$ based on DM from the first CHIME/FRB catalog. The observed DM of FRBs is generally composed of several distinct components \citep{2014ApJ...783L..35D, 2014ApJ...788..189G, 2023MNRAS.519.1823Z}:
    \begin{equation}
    {\rm DM} = {\rm DM_{MW}} + {\rm DM_{Halo}} + {\rm DM_{IGM}} + \frac{{\rm DM_{Host}}}{1+z},
\end{equation}
    where $\rm DM_{MW}, DM_{Halo}, DM_{IGM},$ and $\rm DM_{Host}$ denote the contributions from the Milky Way interstellar medium, the Milky Way halo, the intergalactic medium (IGM), and FRB host galaxy. $\rm DM_{MW}$ is estimated based on the NE2001 model \citep{2002astro.ph..7156C}. Therefore, the extragalactic DM can be defined as
    \begin{equation}
    {\rm DM_{E}} = {\rm DM} - {\rm DM_{MW}} - {\rm DM_{Halo}} = {\rm DM_{IGM}} + \frac{{\rm DM_{Host}}}{1+z}.
\end{equation}
    Following the previous studies, we 
    adopt ${\rm DM_{Halo} }= 30$ $\mathrm{pc} \, \mathrm{cm}^{-3}$ \citep{2015MNRAS.451.4277D, 2021MNRAS.501.5319A, 2023MNRAS.519.1823Z} and ${\rm DM_{Host}} = 130$ $\mathrm{pc} \, \mathrm{cm}^{-3}$ \citep{2024arXiv240916952C}. Note that the $\rm DM_{Host}$ values are provided by \citet{2024arXiv240916952C} recently based on the IllustrisTNG simulation. $\rm DM_{IGM}$ can be expressed in the standard $\Lambda$CDM model as (\citep{2014ApJ...783L..35D, 2014ApJ...788..189G, 2020Natur.581..391M, 2024arXiv240916952C}) :
    \begin{equation}
        \langle{\rm DM_{IGM}}(z)\rangle=\frac{210c\Omega_{\rm b}h_{70}}{8\pi G m_{\rm p}}\int_0^z\frac{f_{\rm d}({z})\,f_{\rm e}({z})\left(1+{z}\right)d{z}}{\sqrt{\Omega_{\Lambda}+\Omega_{\rm m}(1+z)^3}},
    \end{equation}
    where $G$ is the  gravitational constant and $m_{\rm p}$ is the mass of proton. $f_{\rm d}(z)$ denotes the fraction of baryon mass in IGM and $f_{\rm e}(z)$ is the ionized electron number fraction per baryon. In principle, these two parameters are functions of $z$. In this work, we refer to the recent constraints provided by \citet{2024arXiv240916952C} and adopt $f_{\rm d} = 0.93$, $f_{\rm e} = 0.875$ and $\Omega_{\rm b} h_{70} = 0.049$. Other cosmological parameters are adopted from {\tt Planck2018}, with the specific values detailed in the introduction. For FRBs with a small $\rm DM_{E}$, the $\rm DM_{Host}$ may dominate over $\rm DM_{IGM}$, leading to significant uncertainties in redshift estimation. Therefore, we only consider FRBs with $\mathrm{DM_E} > 160 \, \mathrm{pc} \, \mathrm{cm}^{-3}$ when inferring redshifts.

With the values of $z$, $E_{\rm{iso}}$, $L_{\rm{iso}}$, and $T_{\rm{B}}$ of the FRBs can be 
derived. $E_{\rm{iso}}$ and $L_{\rm{iso}}$ is estimated by \citep{2018ApJ...867L..21Z}:

\begin{equation}
\begin{aligned}
E_{\rm{iso}} & \simeq \frac{4 \pi D_{\mathrm{L}}^{2}}{(1+z)} F_{\mathrm{obs}} \nu \\
&=\left(10^{39} \mathrm{erg}\right) \frac{4 \pi}{(1+z)}\left(\frac{D_{\mathrm{L}}}{10^{28} \mathrm{~cm}}\right)^{2} \frac{F_{\mathrm{obs}}}{\mathrm{Jy} \cdot \mathrm{ms}} \frac{\nu}{\mathrm{GHz}},
\end{aligned}
\end{equation}

\begin{equation}
\begin{aligned}
L_{\rm{iso}} & \simeq 4 \pi D_{\mathrm{L}}^{2} S_{\nu, \mathrm{p}} \nu\\
&=\left(10^{42} \operatorname{erg~s}^{-1}\right) 4 \pi\left(\frac{D_{\mathrm{L}}}{10^{28} \mathrm{~cm}}\right)^{2} \frac{S_{\nu, \mathrm{p}}}{\mathrm{Jy}} \frac{\nu}{\mathrm{GHz}},
\end{aligned}
\end{equation}
where $F_{\mathrm{obs}}$ is the time integral fluence (in units of erg~ cm$^{-2}~$Hz$^{-1}$ or Jy$\cdot$ms), $S_{\nu,\mathrm{p}}$ is the peak flux (in units of erg ~s$^{-1}$~cm$^{-2}$~Hz$^{-1}$ or Jy), and $D_{\mathrm{L}}$ is the luminosity distance.

The brightness temperature $T_{\rm{B}}$ can be calculated according to the expression by \citep{2019A&ARv..27....4P,2022A&A...657L...7X}:

\begin{figure*}
\centering
\includegraphics[angle=0,scale=0.60]{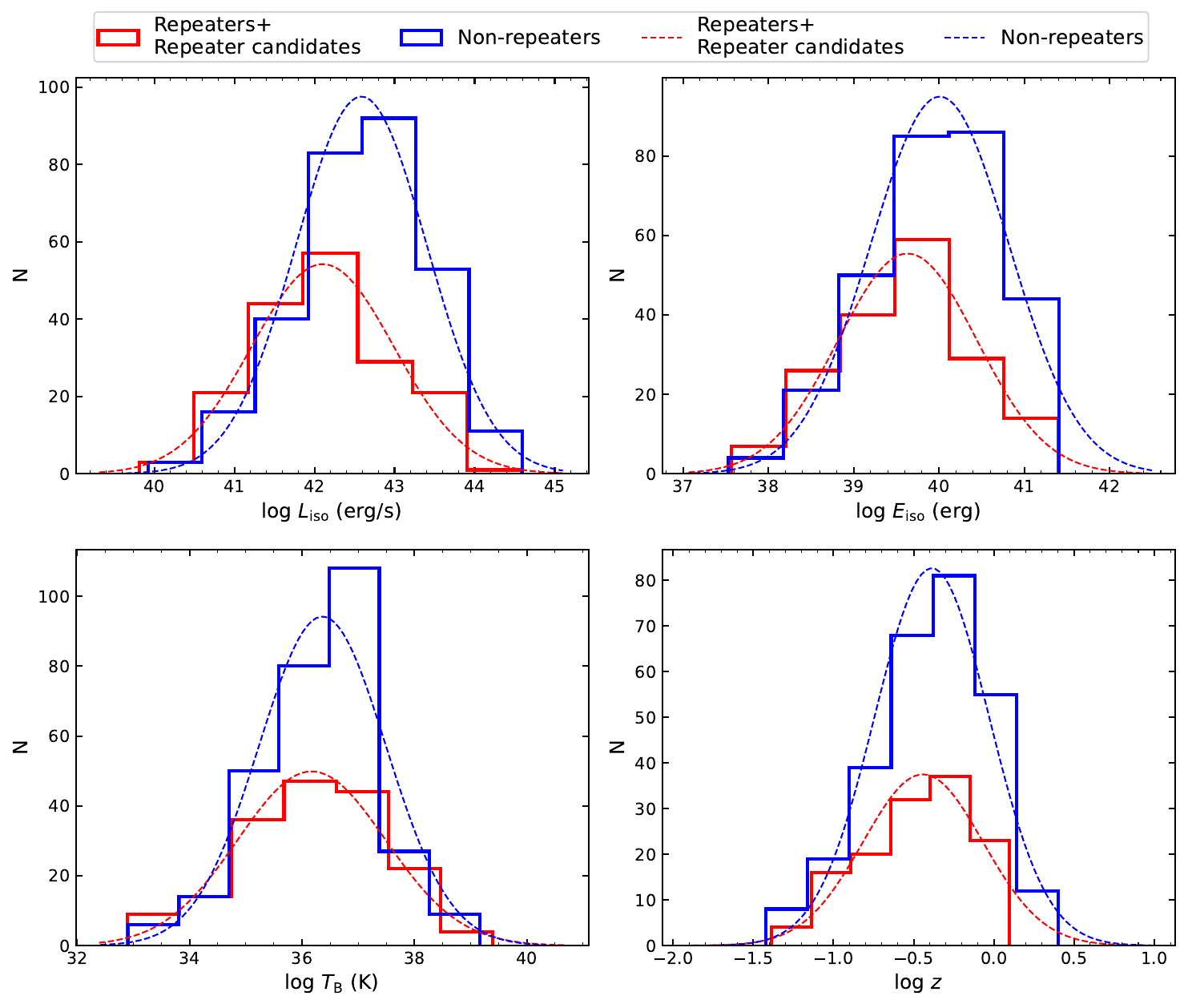}
\caption{Distributions of the luminosity $L_{\rm{iso}}$, isotropic energy $E_{\rm{iso}}$, brightness temperature $T_{\rm{B}}$, and redshift $z$. The red and blue lines denote the repeaters and non-repeaters, respectively. The dotted lines are Gaussian fitting curves.
\label{fig:figure10}}
\end{figure*}

\begin{equation}
\begin{aligned}
T_{\rm{B}} & \simeq \frac{S_{\nu, \mathrm{p}} D_{\mathrm{L}}^{2}}{2 \pi \kappa_{B}(\nu \Delta t)^{2}} \\
& =1.1 \times 10^{35} \mathrm{~K}\left(\frac{S_{\nu, p}}{\mathrm{Jy}}\right)\left(\frac{D_{\mathrm{L}}}{\mathrm{Gpc}}\right)^{2}\left(\frac{\nu}{\mathrm{GHz}}\right)^{-2}\left(\frac{\Delta t}{\mathrm{~ms}}\right)^{-2},
\end{aligned}
\end{equation}
where $\kappa_{B}$ represents boltzmann constant, and $\Delta t$ represents the duration of the burst. 
It is worth noting that in the work of \citet{2016PASA...33...45P} and \citet{2018ApJ...867L..21Z}, they respectively adopted the telescope's bandwidth $B$ and central frequency $\nu _{\mathrm{c}}$ to derive the values of physical parameters such as $E_{\mathrm{iso}}$, $L_{\mathrm{iso}}$, and $T_{\mathrm{B}}$ . However, a significant number of bursts in this sample have completely exceeded CHIME's receive bandwidth. Therefore, it is more appropriate to use the median frequency bandwidth, $\nu$ = 600.0 MHz, rather than the equipment bandwidth. For FRBs that do not exceed the receiving bandwidth, the appropriate value is $\nu$ = 400.0 MHz. 

We examine the parameter distributions of $L_{\mathrm{iso}}$, $E_{\mathrm{iso}}$, $T_{\mathrm{B}}$, and $z$ in the S\textsubscript{grp} sample, as depicted in Figure \ref{fig:figure10}. The repeaters and non-repeaters are shown by the red and blue stepped lines, respectively. 
The distributions of these parameters are consistent and generally follow the log-normal distributions for both repeaters and non-repeaters, but in general, the four parameters of repeaters have a wider distribution range. We find that the $L_\mathrm{iso}$, $E_\mathrm{iso}$, $T_\mathrm{B}$, and $z$ values of repeaters are on average smaller than those of non-repeaters, and with lower minimum. To provide a quantitative result for the S\textsubscript{grp} sample, we perform both Gaussian fitting and the AD test, with the results listed in Table \ref{tab:tab5}. The null hypothesis of the AD test is consistent with Section \ref{ssec:paramt}. According to the statistical results in Table \ref{tab:tab5}, the p-values for $L_{\mathrm{iso}}$, $E_{\mathrm{iso}}$, $T_{\mathrm{B}}$, and $z$ between repeaters and non-repeaters are all smaller than 0.05, indicating that the distributions of these four physical parameters differ between repeaters and non-repeaters. Furthermore, we perform AD tests on the distributions of these parameters after excluding sub-bursts and multiple repeated bursts. 
Notably, the results further support the consistency in the distribution of $T_{\mathrm{B}}$ between repeaters and non-repeaters, with $p_{\rm AD} = 0.25$, suggesting that these two populations may exhibit similar radiation power. However, significant differences persist in $L_{\mathrm{iso}}$ and $E_{\mathrm{iso}}$ between the two populations. This suggests that there may be differences in the emission mechanism \citep{2024NatAs...8..337K}. Although the AD test result for $z$ distribution yields $p_{\rm AD} = 0.114$ , suggesting that the z-distributions of repeaters and non-repeaters are consistent, this is likely due to the exclusion of low-redshift FRBs with  $\mathrm{DM_E} < 160 \, \mathrm{pc} \, \mathrm{cm}^{-3}$ in our calculation. 
This result may be affected by the current sample size limitations or observational selection effects. We will further investigate and verify this phenomenon as larger datasets become available in the future.

\begin{table}\scriptsize
\begin{center}
\caption{The comparison of median values and distributions of FRB intrinsic property parameters.}
\label{tab:tab5}
\tabletypesize{\scriptsize}
\setlength{\tabcolsep}{2pt}
\begin{tabular}{lccc}
\hline

Parameter    & Repeater   & Non-repeater  & $p_{\rm AD}$\\
\hline
$\log_{10} \left(L_{\rm iso}/{\rm erg}\,{\rm s}^{-1}\right)$  & $42.10\pm{0.89}$ & $42.59\pm{0.82}$ & $\leq 0.001$\\
$\log_{10} \left(E_{\rm iso}/{\rm erg}\right)$                & $39.63\pm{0.81}$ & $40.01\pm{0.81}$ & $\leq 0.001$\\
$\log_{10} \left(T_{\rm B}/{\rm K}\right)$                    & $36.17\pm{1.32}$ & $36.36\pm{1.13}$ & $0.069$\\
$\log_{10}z$                                                    & $-0.44\pm{0.37}$  & $-0.39\pm{0.36}$ & 0.114\\
\hline
\end{tabular}
\end{center}
\end{table}

Due to the current limitations of observational data, this work is based solely on the CHIME/FRB sample, covering a specific frequency range. With advancements in observational technology and the accumulation of data, particularly from expanded coverage at lower and higher frequencies \citep{2017ApJ...846L..19P,2021ATel14605....1K,2021Natur.598..267L,2021Natur.596..505P,2023ApJ...957L...8S,2023NatAs...7.1486S}, future research is expected to provide deeper insights into the physical origins and radiation mechanisms of different types of FRBs across a broader frequency spectrum.

\section{Conclusion} \label{sec:concl}
In this work, we apply the unsupervised dimensionality reduction algorithm, t-SNE, to 
recategorize 500 non-repeating and 94 repeating FRBs in the first CHIME/FRB sample. The results are illustrated in Figure \ref{fig:figure8}, where FRBs are divided into two clusters.
We have the following findings.
\begin{enumerate}
    \item By using only seven observed parameters, i.e., $S_{\nu}$, $F_{\nu}$, $\Delta t_\mathrm{WS}$, $\Delta t_\mathrm{ST}$, $\gamma$, $r$ and $\nu_\mathrm{Low}$, we clearly distinguished repeaters
    candidates from the apparent non-repeaters in the first CHIME/FRB sample. 
    The permutation feature importance result shows that the parameters $r$ and $\gamma$, which are related to the spectral morphology of FRBs, are most crucial for classifying repeaters and non-repeaters, while $r$ is particularly influential. 
    \item We identify 163 repeater candidates within the first CHIME/FRB catalog and list them in Table \ref{tab:tab6}. The result demonstrates significant overlap with known repeaters and suggests an increase in the proportion of repeating FRBs from $15.8\%$ to $43.3\%$. 
    It also indicates that many repeating FRBs may still be undetected or were previously overlooked.
    \item We validate our predictions with 25 new repeating FRB sources from the CHIME/FRB catalog.  Recent observations confirmed six formerly non-repeating FRBs as repeaters \citep{2023ApJ...947...83C}. Our method successfully predicted five of these six FRBs as repeaters. Additionally, during validation and dimensionality reduction, we identified two specific repeaters that were misclassified. These repeaters exhibit characteristics that differ from typical repeaters and more closely resemble non-repeaters. Currently, there is no definitive explanation for this discrepancy, and further continuous observation of larger samples is needed to verify and elucidate this phenomenon.
    \item We confirm that the two spectral morphology parameters, i.e. $r$ and $\gamma$, 
    significantly differ between repeaters and non-repeaters. 
    Most of the repeaters' frequency spectra are narrowband, while the non-repeaters' 
    are broadband.
    This suggests that repeaters and non-repeaters may originate from different radiation mechanisms or environments. After reclassifying the first CHIME/FRB catalog, 
    the repeaters (including repeater candidates) exhibit lower values of $L_{\rm{iso}}$, $E_{\rm{iso}}$, $T_{\rm{B}}$, and $z$ compared to non-repeaters. The AD test results also indicate that the distributions of $L_{\rm{iso}}$ and $E_{\rm{iso}}$ differ between repeaters and non-repeaters, while the $T_{\mathrm{B}}$ distributions are similar.
\end{enumerate}

This work highlights that the machine learning method can overcome inefficiencies in traditional observation techniques, revealing hidden repeaters and aiding understanding of their fundamental physical differences. 
A list of 163 repeater candidates identified from the first CHIME/FRB catalog 
requires validation through further observations.

\section*{Acknowledgments}
We are very grateful to Ze-Wei Zhao, Shu-Lei Ni, Guo-Hong Du, and Si-Yuan Zhu for helpful discussions. We thank the referee and the editor for the useful comments that help significantly improve the paper. 
We acknowledge the support of the National SKA Program of China 
(Grant Nos. 2022SKA0110200, 2022SKA0110203), 
the National Natural Science Foundation of China (Grant Nos. 12473001, 11975072, 11835009), and the 111 Project (Grant No. B16009).

\section*{Data Availability}
The 163 repeater candidates identified from the first CHIME/FRB catalog are listed in Table \ref{tab:tab6} in APPENDIX.

\bibliography{FRB_tSNE}

\section*{APPENDIX}
\label{app:appendix}

This work uses unsupervised machine learning methods to identify 163 repeater candidates in the first CHIME/FRB catalog, with their observational parameters listed in Table \ref{tab:tab6}. These candidates exhibit characteristics highly similar to repeaters, suggesting that they are more likely to repeat. Thus, they can be considered key observational targets for future studies on repeaters. Table \ref{tab:tab6} presents their positional information, including Right Ascension (R.A.), declination (Decl.), and DM, along with seven selected feature parameters describing the intrinsic or environmental properties of FRBs, i.e., $S_{\nu}$, $F_{\nu}$, $\Delta t_\mathrm{WS}$, $\Delta t_\mathrm{ST}$, $\gamma$, $r$ and $\nu_\mathrm{Low}$.

\startlongtable
\begin{deluxetable*}{lccccccccccc}
\tablecaption{The repeater candidates identified by unsupervised machine learning methods. The `Sub\_num' column represents the sub-burst number of FRBs from the first CHIME/FRB catalog. A detailed description of these properties is provided in Section \ref{sec:data}.\label{tab:tab6}}
\tablewidth{700pt}
\tabletypesize{\scriptsize}
\tablehead{
\colhead{Name} & \colhead{Sub\_num} & \colhead{R.A. (J2000)} & \colhead{Decl. (J2000)} & \colhead{DM}& \colhead{$S_{\nu}$\vspace{-2pt}} & \colhead{$F_{\nu}$\vspace{-2pt}} & \colhead{$\Delta t_\mathrm{WS}$} & \colhead{$\Delta t_\mathrm{ST}$} & \colhead{$\gamma$} & \colhead{$r$} & \colhead{$\nu_\mathrm{Low}$\vspace{-2pt}}  \\
\colhead{} & \colhead{} & \colhead{(\(^\circ\)\vspace{-2pt})} & \colhead{(\(^\circ\)\vspace{-2pt})}  & \colhead{(pc cm$^{-3}$)}& \colhead{(Jy)} & \colhead{(Jy ms)} & \colhead{(ms)} & \colhead{(ms)} & \colhead{} & \colhead{} & \colhead{(MHz)}  \\[-14pt]
}

\startdata
FRB20180725A	&	0	&	93.42	$\pm$	0.039 	&	67.07	$\pm$	0.210 	&	715.809 	&	1.70 	&	4.10 	&	0.296 	&	1.100 	&	$	38.20 	$	&	$	-45.80 	$	&	400.2	\\
FRB20180729A	&	0	&	199.4	$\pm$	0.120 	&	55.58	$\pm$	0.084 	&	109.594 	&	11.70 	&	17.00 	&	0.100 	&	0.157 	&	$	16.46 	$	&	$	-30.21 	$	&	400.2	\\
FRB20180729B	&	0	&	89.93	$\pm$	0.270 	&	56.5	$\pm$	0.240 	&	317.224 	&	0.92 	&	1.20 	&	0.314 	&	0.660 	&	$	14.50 	$	&	$	-14.60 	$	&	400.2	\\
FRB20180801A	&	0	&	322.53	$\pm$	0.059 	&	72.72	$\pm$	0.220 	&	655.728 	&	1.11 	&	7.90 	&	0.580 	&	5.540 	&	$	60.00 	$	&	$	-75.50 	$	&	400.2	\\
FRB20180904A	&	0	&	286.58	$\pm$	0.170 	&	81.22	$\pm$	0.120 	&	361.137 	&	3.80 	&	6.00 	&	0.528 	&	0.550 	&	$	12.12 	$	&	$	-23.18 	$	&	400.2	\\
FRB20180916C	&	0	&	107.15	$\pm$	0.230 	&	45.08	$\pm$	0.240 	&	2252.873 	&	0.39 	&	2.10 	&	4.060 	&	5.100 	&	$	47.00 	$	&	$	-50.00 	$	&	400.2	\\
FRB20180918A	&	0	&	301.27	$\pm$	0.220 	&	64.96	$\pm$	0.200 	&	1453.988 	&	1.45 	&	4.10 	&	0.550 	&	1.460 	&	$	13.40 	$	&	$	-15.50 	$	&	400.2	\\
FRB20180920A	&	0	&	78.89	$\pm$	0.210 	&	28.29	$\pm$	0.230 	&	555.660 	&	0.86 	&	8.50 	&	2.220 	&	9.100 	&	$	20.10 	$	&	$	-26.30 	$	&	400.2	\\
FRB20180920B	&	0	&	191.09	$\pm$	0.230 	&	63.52	$\pm$	0.240 	&	463.400 	&	0.35 	&	1.70 	&	2.330 	&	1.730 	&	$	12.30 	$	&	$	-121.00 	$	&	400.2	\\
FRB20180923A	&	0	&	327.61	$\pm$	0.038 	&	71.92	$\pm$	0.200 	&	219.440 	&	0.76 	&	1.20 	&	0.150 	&	0.211 	&	$	18.20 	$	&	$	-57.40 	$	&	400.2	\\
FRB20180923D	&	0	&	169.08	$\pm$	0.020 	&	48.75	$\pm$	0.070 	&	329.400 	&	2.40 	&	2.20 	&	0.100 	&	0.114 	&	$	20.90 	$	&	$	-92.70 	$	&	400.2	\\
FRB20180925B	&	0	&	145.45	$\pm$	0.210 	&	20.99	$\pm$	0.089 	&	667.866 	&	0.76 	&	2.70 	&	1.150 	&	1.400 	&	$	15.00 	$	&	$	-41.50 	$	&	400.2	\\
FRB20180928A	&	0	&	312.95	$\pm$	0.040 	&	30.85	$\pm$	0.053 	&	252.768 	&	1.34 	&	2.50 	&	0.269 	&	0.148 	&	$	-2.93 	$	&	$	-39.70 	$	&	400.2	\\
FRB20181012B	&	0	&	206.33	$\pm$	0.210 	&	64.15	$\pm$	0.065 	&	715.189 	&	0.49 	&	1.44 	&	0.560 	&	0.260 	&	$	20.90 	$	&	$	-154.00 	$	&	400.2	\\
FRB20181014A	&	0	&	46.01	$\pm$	0.220 	&	63.33	$\pm$	0.200 	&	1314.890 	&	0.99 	&	2.70 	&	2.050 	&	2.300 	&	$	3.70 	$	&	$	0.50 	$	&	400.2	\\
FRB20181014C	&	0	&	117.87	$\pm$	0.220 	&	41.59	$\pm$	0.230 	&	752.167 	&	0.57 	&	1.48 	&	0.790 	&	1.000 	&	$	18.00 	$	&	$	-30.50 	$	&	449.9	\\
FRB20181017B	&	0	&	237.76	$\pm$	0.230 	&	78.5	$\pm$	0.250 	&	307.369 	&	1.06 	&	6.50 	&	2.310 	&	4.300 	&	$	61.00 	$	&	$	-77.00 	$	&	400.2	\\
FRB20181022D	&	0	&	179.18	$\pm$	0.160 	&	36.53	$\pm$	0.160 	&	514.334 	&	2.90 	&	6.20 	&	0.594 	&	0.630 	&	$	18.20 	$	&	$	-14.40 	$	&	400.2	\\
FRB20181022E	&	0	&	221.18	$\pm$	0.210 	&	27.13	$\pm$	0.220 	&	285.986 	&	0.69 	&	2.08 	&	0.400 	&	0.632 	&	$	8.70 	$	&	$	-42.30 	$	&	400.2	\\
FRB20181027A	&	0	&	131.9	$\pm$	0.220 	&	-4.24	$\pm$	0.340 	&	727.744 	&	4.90 	&	22.00 	&	0.720 	&	4.720 	&	$	-0.40 	$	&	$	6.10 	$	&	400.2	\\
FRB20181030C	&	0	&	309.83	$\pm$	0.220 	&	3.99	$\pm$	0.300 	&	668.760 	&	1.60 	&	5.50 	&	0.610 	&	0.740 	&	$	10.00 	$	&	$	-22.20 	$	&	400.2	\\
FRB20181030E	&	0	&	135.67	$\pm$	0.012 	&	8.89	$\pm$	0.190 	&	159.690 	&	2.00 	&	6.30 	&	0.400 	&	0.973 	&	$	22.30 	$	&	$	-69.00 	$	&	400.2	\\
FRB20181101A	&	0	&	21.26	$\pm$	0.011 	&	53.88	$\pm$	0.160 	&	1472.678 	&	0.50 	&	10.70 	&	6.030 	&	10.000 	&	$	16.40 	$	&	$	-37.70 	$	&	551	\\
FRB20181115A	&	0	&	142.98	$\pm$	0.039 	&	56.4	$\pm$	0.180 	&	981.613 	&	0.44 	&	1.92 	&	1.830 	&	2.100 	&	$	19.60 	$	&	$	-62.00 	$	&	401.7	\\
FRB20181117B	&	1	&	81.09	$\pm$	0.190 	&	79.99	$\pm$	0.099 	&	538.200 	&	3.60 	&	11.00 	&	1.280 	&	1.300 	&	$	11.40 	$	&	$	-31.20 	$	&	400.2	\\
FRB20181117C	&	0	&	53.21	$\pm$	0.170 	&	25.73	$\pm$	0.180 	&	1773.739 	&	1.57 	&	3.00 	&	0.100 	&	2.370 	&	$	48.60 	$	&	$	-46.30 	$	&	400.2	\\
FRB20181123A	&	0	&	300.76	$\pm$	0.023 	&	55.87	$\pm$	0.180 	&	798.718 	&	0.99 	&	2.50 	&	0.580 	&	1.920 	&	$	25.60 	$	&	$	-64.30 	$	&	400.2	\\
FRB20181125A	&	0	&	147.94	$\pm$	0.180 	&	33.93	$\pm$	0.041 	&	272.190 	&	0.39 	&	3.20 	&	1.280 	&	1.500 	&	$	9.00 	$	&	$	-54.00 	$	&	400.2	\\
FRB20181125A	&	1	&	147.94	$\pm$	0.180 	&	33.93	$\pm$	0.041 	&	272.190 	&	0.39 	&	3.20 	&	1.440 	&	1.500 	&	$	7.30 	$	&	$	-41.80 	$	&	400.2	\\
FRB20181125A	&	2	&	147.94	$\pm$	0.180 	&	33.93	$\pm$	0.041 	&	272.190 	&	0.39 	&	3.20 	&	1.580 	&	1.500 	&	$	7.30 	$	&	$	-58.00 	$	&	436.9	\\
FRB20181126A	&	0	&	262.05	$\pm$	0.180 	&	81.17	$\pm$	0.190 	&	494.217 	&	3.50 	&	9.40 	&	0.407 	&	0.165 	&	$	17.93 	$	&	$	-90.70 	$	&	400.2	\\
FRB20181127A	&	0	&	243.8	$\pm$	0.230 	&	25.43	$\pm$	0.250 	&	930.317 	&	0.78 	&	2.90 	&	0.740 	&	0.582 	&	$	18.50 	$	&	$	-51.30 	$	&	400.2	\\
FRB20181128C	&	0	&	268.77	$\pm$	0.023 	&	49.71	$\pm$	0.200 	&	618.350 	&	0.39 	&	3.40 	&	2.300 	&	2.320 	&	$	27.40 	$	&	$	-75.00 	$	&	442.1	\\
FRB20181128C	&	1	&	268.77	$\pm$	0.023 	&	49.71	$\pm$	0.200 	&	618.350 	&	0.39 	&	3.40 	&	0.100 	&	2.320 	&	$	23.30 	$	&	$	-60.00 	$	&	400.2	\\
FRB20181129B	&	0	&	307.56	$\pm$	0.260 	&	81.32	$\pm$	0.340 	&	405.905 	&	4.00 	&	9.50 	&	0.364 	&	0.830 	&	$	73.80 	$	&	$	-112.00 	$	&	400.2	\\
FRB20181130A	&	0	&	355.19	$\pm$	0.031 	&	46.49	$\pm$	0.170 	&	220.090 	&	0.97 	&	1.27 	&	0.480 	&	0.510 	&	$	2.70 	$	&	$	-45.70 	$	&	400.2	\\
FRB20181203A	&	0	&	33.6	$\pm$	0.027 	&	23.57	$\pm$	0.190 	&	635.926 	&	1.74 	&	3.60 	&	0.171 	&	1.440 	&	$	13.20 	$	&	$	-8.50 	$	&	400.2	\\
FRB20181203B	&	0	&	47.31	$\pm$	0.004 	&	24.02	$\pm$	0.120 	&	375.387 	&	1.45 	&	4.50 	&	0.578 	&	2.320 	&	$	47.80 	$	&	$	-43.60 	$	&	400.2	\\
FRB20181209A	&	0	&	98.16	$\pm$	0.210 	&	68.69	$\pm$	0.210 	&	328.656 	&	2.50 	&	3.20 	&	0.597 	&	0.640 	&	$	0.42 	$	&	$	-25.50 	$	&	400.2	\\
FRB20181213B	&	0	&	183.52	$\pm$	0.220 	&	53.7	$\pm$	0.230 	&	626.593 	&	0.75 	&	1.70 	&	0.850 	&	1.200 	&	$	45.60 	$	&	$	-45.10 	$	&	400.2	\\
FRB20181214A	&	0	&	70	$\pm$	0.180 	&	43.07	$\pm$	0.180 	&	468.148 	&	0.16 	&	0.41 	&	0.533 	&	0.442 	&	$	23.30 	$	&	$	-139.00 	$	&	400.2	\\
FRB20181214D	&	0	&	178.57	$\pm$	0.200 	&	46.71	$\pm$	0.190 	&	1177.319 	&	0.55 	&	9.00 	&	0.990 	&	18.000 	&	$	12.20 	$	&	$	-10.90 	$	&	400.2	\\
FRB20181214F	&	0	&	252.62	$\pm$	0.047 	&	32.44	$\pm$	0.220 	&	2105.760 	&	0.31 	&	2.21 	&	2.300 	&	1.520 	&	$	8.00 	$	&	$	-65.00 	$	&	403.2	\\
FRB20181216A	&	0	&	306.28	$\pm$	0.220 	&	53.53	$\pm$	0.230 	&	542.738 	&	0.94 	&	1.70 	&	0.220 	&	0.890 	&	$	14.80 	$	&	$	-16.10 	$	&	400.2	\\
FRB20181218A	&	0	&	5.06	$\pm$	0.190 	&	71.35	$\pm$	0.076 	&	1874.406 	&	0.83 	&	1.59 	&	1.390 	&	0.221 	&	$	19.00 	$	&	$	-83.60 	$	&	400.2	\\
FRB20181219B	&	0	&	180.79	$\pm$	0.190 	&	71.55	$\pm$	0.350 	&	1952.166 	&	4.60 	&	27.00 	&	0.574 	&	5.260 	&	$	-4.20 	$	&	$	10.30 	$	&	400.2	\\
FRB20181221A	&	0	&	230.58	$\pm$	0.200 	&	25.86	$\pm$	0.210 	&	316.237 	&	1.25 	&	5.80 	&	0.754 	&	1.323 	&	$	62.10 	$	&	$	-128.00 	$	&	400.2	\\
FRB20181221B	&	0	&	306.31	$\pm$	0.004 	&	80.98	$\pm$	0.028 	&	1395.021 	&	0.97 	&	3.30 	&	1.037 	&	1.100 	&	$	25.30 	$	&	$	-61.20 	$	&	400.2	\\
FRB20181222D	&	0	&	188.2	$\pm$	0.230 	&	56.16	$\pm$	0.084 	&	1417.110 	&	0.22 	&	1.23 	&	3.750 	&	1.950 	&	$	8.40 	$	&	$	-41.10 	$	&	400.2	\\
FRB20181222E	&	0	&	50.64	$\pm$	0.044 	&	86.97	$\pm$	0.250 	&	327.980 	&	1.12 	&	5.50 	&	0.254 	&	0.790 	&	$	5.13 	$	&	$	-19.90 	$	&	400.2	\\
FRB20181222E	&	1	&	50.64	$\pm$	0.044 	&	86.97	$\pm$	0.250 	&	327.980 	&	1.12 	&	5.50 	&	0.910 	&	0.790 	&	$	9.90 	$	&	$	-23.80 	$	&	400.2	\\
FRB20181223B	&	0	&	174.89	$\pm$	0.220 	&	21.59	$\pm$	0.240 	&	565.655 	&	0.68 	&	4.10 	&	1.570 	&	3.500 	&	$	33.30 	$	&	$	-41.00 	$	&	475.4	\\
FRB20181228B	&	0	&	250.43	$\pm$	0.210 	&	63.85	$\pm$	0.210 	&	568.651 	&	0.40 	&	1.67 	&	0.100 	&	1.159 	&	$	59.30 	$	&	$	-353.00 	$	&	506.3	\\
FRB20181229B	&	0	&	238.37	$\pm$	0.230 	&	19.78	$\pm$	0.260 	&	389.047 	&	0.42 	&	4.90 	&	3.360 	&	5.100 	&	$	22.00 	$	&	$	-103.00 	$	&	400.2	\\
FRB20181230A	&	0	&	346.69	$\pm$	0.220 	&	83.37	$\pm$	0.260 	&	769.611 	&	0.94 	&	18.00 	&	1.640 	&	41.000 	&	$	33.00 	$	&	$	-27.80 	$	&	400.2	\\
FRB20181231B	&	0	&	128.77	$\pm$	0.200 	&	55.99	$\pm$	0.180 	&	197.170 	&	0.89 	&	2.34 	&	0.337 	&	1.750 	&	$	59.60 	$	&	$	-60.00 	$	&	400.2	\\
FRB20190101B	&	0	&	307.77	$\pm$	0.230 	&	29.89	$\pm$	0.230 	&	1323.906 	&	1.02 	&	4.40 	&	0.320 	&	5.160 	&	$	41.70 	$	&	$	-36.10 	$	&	400.2	\\
FRB20190102A	&	0	&	9.26	$\pm$	0.180 	&	26.72	$\pm$	0.057 	&	699.173 	&	1.12 	&	4.20 	&	0.824 	&	0.986 	&	$	28.90 	$	&	$	-67.80 	$	&	400.2	\\
FRB20190106A	&	0	&	22.19	$\pm$	0.240 	&	46.12	$\pm$	0.250 	&	340.061 	&	0.27 	&	0.81 	&	0.940 	&	5.800 	&	$	9.00 	$	&	$	-5.00 	$	&	400.2	\\
FRB20190106B	&	0	&	335.63	$\pm$	0.180 	&	46.13	$\pm$	0.180 	&	316.594 	&	1.70 	&	3.80 	&	0.578 	&	0.600 	&	$	16.58 	$	&	$	-68.00 	$	&	400.2	\\
FRB20190109B	&	0	&	253.47	$\pm$	0.230 	&	1.25	$\pm$	0.330 	&	175.168 	&	1.20 	&	3.00 	&	0.340 	&	0.261 	&	$	2.50 	$	&	$	-65.00 	$	&	400.2	\\
FRB20190110A	&	0	&	64.95	$\pm$	0.033 	&	47.44	$\pm$	0.086 	&	472.753 	&	1.54 	&	3.80 	&	0.203 	&	0.381 	&	$	6.30 	$	&	$	-118.00 	$	&	400.2	\\
FRB20190112A	&	0	&	257.98	$\pm$	0.015 	&	61.2	$\pm$	0.026 	&	425.847 	&	1.40 	&	16.20 	&	1.640 	&	11.010 	&	$	57.10 	$	&	$	-51.40 	$	&	400.2	\\
FRB20190114A	&	0	&	8.95	$\pm$	0.230 	&	19.17	$\pm$	0.260 	&	887.392 	&	0.55 	&	2.30 	&	1.340 	&	0.380 	&	$	11.10 	$	&	$	-89.00 	$	&	446.1	\\
FRB20190118A	&	0	&	253.31	$\pm$	0.029 	&	11.55	$\pm$	0.084 	&	225.108 	&	9.30 	&	18.00 	&	0.140 	&	0.282 	&	$	19.42 	$	&	$	-56.71 	$	&	400.2	\\
FRB20190118B	&	0	&	39.71	$\pm$	0.250 	&	23.57	$\pm$	0.270 	&	670.890 	&	0.31 	&	3.62 	&	3.700 	&	19.400 	&	$	13.80 	$	&	$	-14.40 	$	&	400.2	\\
FRB20190125A	&	0	&	45.73	$\pm$	0.240 	&	27.81	$\pm$	0.260 	&	564.701 	&	0.37 	&	2.60 	&	3.210 	&	4.100 	&	$	36.00 	$	&	$	-37.00 	$	&	400.2	\\
FRB20190128C	&	0	&	69.8	$\pm$	0.230 	&	78.94	$\pm$	0.380 	&	310.622 	&	0.71 	&	5.90 	&	6.160 	&	7.600 	&	$	22.60 	$	&	$	-55.00 	$	&	400.2	\\
FRB20190129A	&	0	&	45.06	$\pm$	0.210 	&	21.42	$\pm$	0.230 	&	484.761 	&	0.49 	&	5.00 	&	1.130 	&	10.200 	&	$	43.00 	$	&	$	-37.80 	$	&	514.4	\\
FRB20190130A	&	0	&	25.64	$\pm$	0.240 	&	13.16	$\pm$	0.300 	&	1367.461 	&	0.47 	&	4.40 	&	0.990 	&	3.200 	&	$	19.30 	$	&	$	-62.00 	$	&	455.6	\\
FRB20190130B	&	0	&	172.11	$\pm$	0.160 	&	16.05	$\pm$	0.072 	&	989.031 	&	0.77 	&	2.95 	&	0.265 	&	0.769 	&	$	55.40 	$	&	$	-140.80 	$	&	403.3	\\
FRB20190131B	&	0	&	354.72	$\pm$	0.240 	&	11.71	$\pm$	0.230 	&	1805.729 	&	0.99 	&	3.30 	&	0.920 	&	1.100 	&	$	16.50 	$	&	$	-11.90 	$	&	400.2	\\
FRB20190131E	&	0	&	195.65	$\pm$	0.044 	&	80.92	$\pm$	0.270 	&	279.801 	&	3.00 	&	5.10 	&	0.230 	&	0.164 	&	$	22.00 	$	&	$	-63.10 	$	&	400.2	\\
FRB20190203A	&	0	&	133.68	$\pm$	0.170 	&	70.82	$\pm$	0.190 	&	420.573 	&	1.21 	&	4.00 	&	0.550 	&	0.832 	&	$	25.00 	$	&	$	-75.00 	$	&	400.2	\\
FRB20190204A	&	0	&	161.33	$\pm$	0.230 	&	61.53	$\pm$	0.240 	&	449.639 	&	0.24 	&	1.50 	&	1.810 	&	0.880 	&	$	2.40 	$	&	$	-28.00 	$	&	400.2	\\
FRB20190205A	&	0	&	342.22	$\pm$	0.250 	&	83.37	$\pm$	0.300 	&	695.389 	&	0.74 	&	1.70 	&	0.602 	&	0.690 	&	$	18.30 	$	&	$	-47.30 	$	&	400.2	\\
FRB20190206A	&	0	&	244.85	$\pm$	0.220 	&	9.36	$\pm$	0.260 	&	188.336 	&	1.40 	&	9.10 	&	0.804 	&	2.740 	&	$	38.00 	$	&	$	-65.70 	$	&	400.2	\\
FRB20190206B	&	0	&	49.76	$\pm$	0.250 	&	79.5	$\pm$	0.390 	&	352.520 	&	0.95 	&	9.60 	&	7.100 	&	9.000 	&	$	11.60 	$	&	$	-24.60 	$	&	400.2	\\
FRB20190208C	&	0	&	141.55	$\pm$	0.037 	&	83.56	$\pm$	0.220 	&	238.392 	&	1.27 	&	1.74 	&	0.411 	&	0.450 	&	$	18.70 	$	&	$	-54.60 	$	&	400.2	\\
FRB20190210D	&	0	&	307.8	$\pm$	0.082 	&	55.46	$\pm$	0.180 	&	359.148 	&	1.37 	&	2.50 	&	0.580 	&	0.273 	&	$	20.70 	$	&	$	-24.40 	$	&	400.2	\\
FRB20190210E	&	0	&	313.65	$\pm$	0.280 	&	86.67	$\pm$	0.310 	&	580.580 	&	0.69 	&	1.45 	&	0.960 	&	1.100 	&	$	13.40 	$	&	$	-40.50 	$	&	400.2	\\
FRB20190211A	&	0	&	67.06	$\pm$	0.190 	&	68.64	$\pm$	0.200 	&	1188.256 	&	1.47 	&	5.80 	&	0.360 	&	3.290 	&	$	38.90 	$	&	$	-39.30 	$	&	400.2	\\
FRB20190213D	&	0	&	336.45	$\pm$	0.021 	&	52.71	$\pm$	0.140 	&	1346.848 	&	1.00 	&	2.20 	&	0.626 	&	0.700 	&	$	26.20 	$	&	$	-25.30 	$	&	400.2	\\
FRB20190218B	&	0	&	268.7	$\pm$	0.220 	&	17.93	$\pm$	0.260 	&	547.868 	&	0.57 	&	5.90 	&	2.050 	&	14.100 	&	$	46.20 	$	&	$	-60.00 	$	&	564.6	\\
FRB20190221A	&	0	&	132.6	$\pm$	0.050 	&	9.9	$\pm$	0.260 	&	223.806 	&	1.23 	&	2.33 	&	0.970 	&	0.408 	&	$	5.10 	$	&	$	-23.90 	$	&	400.2	\\
FRB20190221C	&	0	&	316.15	$\pm$	0.200 	&	54.67	$\pm$	0.190 	&	2042.300 	&	0.59 	&	7.10 	&	1.180 	&	13.100 	&	$	16.20 	$	&	$	-16.00 	$	&	526.2	\\
FRB20190222C	&	0	&	239.18	$\pm$	0.040 	&	40.03	$\pm$	0.190 	&	524.007 	&	0.44 	&	0.83 	&	0.676 	&	0.740 	&	$	9.30 	$	&	$	-53.90 	$	&	529.6	\\
FRB20190223A	&	0	&	64.72	$\pm$	0.300 	&	87.65	$\pm$	0.320 	&	389.237 	&	0.47 	&	1.58 	&	0.763 	&	0.870 	&	$	21.80 	$	&	$	-103.00 	$	&	528.7	\\
FRB20190224A	&	0	&	60.53	$\pm$	0.250 	&	83.39	$\pm$	0.290 	&	818.400 	&	0.63 	&	8.50 	&	2.040 	&	5.770 	&	$	2.60 	$	&	$	-60.00 	$	&	527	\\
FRB20190226C	&	0	&	17.46	$\pm$	0.210 	&	26.76	$\pm$	0.056 	&	827.770 	&	0.39 	&	1.41 	&	1.310 	&	1.500 	&	$	6.60 	$	&	$	-42.00 	$	&	408.6	\\
FRB20190228A	&	0	&	183.48	$\pm$	0.005 	&	22.9	$\pm$	0.120 	&	419.083 	&	1.79 	&	35.80 	&	2.250 	&	18.910 	&	$	52.60 	$	&	$	-51.90 	$	&	400.2	\\
FRB20190301D	&	0	&	278.72	$\pm$	0.220 	&	74.68	$\pm$	0.093 	&	1160.692 	&	0.39 	&	1.50 	&	0.360 	&	0.535 	&	$	3.60 	$	&	$	-29.40 	$	&	483.4	\\
FRB20190304A	&	0	&	124.51	$\pm$	0.036 	&	74.61	$\pm$	0.210 	&	483.727 	&	0.71 	&	2.90 	&	0.408 	&	0.901 	&	$	6.30 	$	&	$	-67.30 	$	&	400.2	\\
FRB20190304C	&	0	&	223.01	$\pm$	0.230 	&	26.72	$\pm$	0.250 	&	564.991 	&	0.53 	&	1.32 	&	0.948 	&	1.100 	&	$	22.30 	$	&	$	-87.00 	$	&	400.2	\\
FRB20190308B	&	0	&	38.59	$\pm$	0.040 	&	83.62	$\pm$	0.300 	&	180.180 	&	1.11 	&	1.39 	&	0.186 	&	0.134 	&	$	18.60 	$	&	$	-52.90 	$	&	400.2	\\
FRB20190308B	&	1	&	38.59	$\pm$	0.040 	&	83.62	$\pm$	0.300 	&	180.180 	&	1.11 	&	1.39 	&	0.520 	&	0.134 	&	$	9.50 	$	&	$	-37.00 	$	&	400.2	\\
FRB20190308C	&	0	&	188.36	$\pm$	0.026 	&	44.39	$\pm$	0.170 	&	500.519 	&	0.47 	&	4.80 	&	0.400 	&	2.290 	&	$	15.20 	$	&	$	-61.00 	$	&	402.2	\\
FRB20190308C	&	1	&	188.36	$\pm$	0.026 	&	44.39	$\pm$	0.170 	&	500.519 	&	0.47 	&	4.80 	&	0.550 	&	2.290 	&	$	13.90 	$	&	$	-60.50 	$	&	409.4	\\
FRB20190309A	&	0	&	278.96	$\pm$	0.230 	&	52.41	$\pm$	0.240 	&	356.900 	&	0.39 	&	0.72 	&	0.581 	&	0.750 	&	$	12.90 	$	&	$	-64.00 	$	&	400.2	\\
FRB20190329A	&	0	&	65.54	$\pm$	0.190 	&	73.63	$\pm$	0.270 	&	188.606 	&	0.52 	&	2.24 	&	1.040 	&	0.900 	&	$	42.00 	$	&	$	-272.00 	$	&	500.2	\\
FRB20190403E	&	0	&	220.22	$\pm$	0.086 	&	86.54	$\pm$	0.270 	&	226.198 	&	3.90 	&	76.00 	&	2.200 	&	18.200 	&	$	31.70 	$	&	$	-36.20 	$	&	400.2	\\
FRB20190403G	&	0	&	81.74	$\pm$	0.220 	&	25.78	$\pm$	0.250 	&	865.311 	&	0.75 	&	1.59 	&	1.590 	&	1.900 	&	$	35.70 	$	&	$	-76.00 	$	&	400.2	\\
FRB20190408A	&	0	&	262.2	$\pm$	0.230 	&	71.6	$\pm$	0.260 	&	863.380 	&	0.64 	&	1.51 	&	0.839 	&	1.000 	&	$	35.70 	$	&	$	-49.00 	$	&	400.2	\\
FRB20190409B	&	0	&	126.65	$\pm$	0.220 	&	63.47	$\pm$	0.210 	&	285.633 	&	0.39 	&	6.80 	&	2.340 	&	20.900 	&	$	21.10 	$	&	$	-34.10 	$	&	454.9	\\
FRB20190410A	&	0	&	263.47	$\pm$	0.230 	&	-2.37	$\pm$	0.380 	&	284.020 	&	1.59 	&	5.80 	&	1.010 	&	1.200 	&	$	43.00 	$	&	$	-85.00 	$	&	400.2	\\
FRB20190410B	&	0	&	265.76	$\pm$	0.020 	&	15.17	$\pm$	0.200 	&	642.170 	&	0.22 	&	0.45 	&	0.423 	&	0.211 	&	$	18.70 	$	&	$	-73.40 	$	&	493.5	\\
FRB20190411C	&	0	&	9.33	$\pm$	0.054 	&	20.5	$\pm$	0.210 	&	233.660 	&	3.19 	&	9.30 	&	1.023 	&	1.100 	&	$	24.30 	$	&	$	-26.10 	$	&	400.2	\\
FRB20190411C	&	1	&	9.33	$\pm$	0.054 	&	20.5	$\pm$	0.210 	&	233.660 	&	3.19 	&	9.30 	&	0.890 	&	1.100 	&	$	11.70 	$	&	$	-12.50 	$	&	400.2	\\
FRB20190415C	&	0	&	74.81	$\pm$	0.230 	&	34.8	$\pm$	0.250 	&	650.182 	&	0.46 	&	0.77 	&	0.550 	&	0.870 	&	$	13.40 	$	&	$	-32.00 	$	&	400.2	\\
FRB20190416B	&	0	&	172.19	$\pm$	0.180 	&	35.95	$\pm$	0.034 	&	575.360 	&	0.69 	&	1.47 	&	0.790 	&	0.491 	&	$	-3.60 	$	&	$	-23.50 	$	&	400.2	\\
FRB20190417C	&	0	&	45.68	$\pm$	0.030 	&	71.26	$\pm$	0.046 	&	320.232 	&	7.90 	&	10.80 	&	0.413 	&	0.430 	&	$	24.78 	$	&	$	-32.41 	$	&	400.2	\\
FRB20190419A	&	0	&	104.98	$\pm$	0.210 	&	64.88	$\pm$	0.071 	&	439.972 	&	0.41 	&	0.77 	&	1.850 	&	2.400 	&	$	1.00 	$	&	$	-29.00 	$	&	400.2	\\
FRB20190420A	&	0	&	106.55	$\pm$	0.200 	&	55.96	$\pm$	0.210 	&	609.100 	&	0.88 	&	3.11 	&	0.770 	&	2.640 	&	$	11.50 	$	&	$	-6.10 	$	&	400.2	\\
FRB20190422A	&	0	&	48.56	$\pm$	0.200 	&	35.15	$\pm$	0.200 	&	452.302 	&	0.60 	&	9.10 	&	3.220 	&	2.700 	&	$	42.00 	$	&	$	-46.90 	$	&	515.2	\\
FRB20190422A	&	1	&	48.56	$\pm$	0.200 	&	35.15	$\pm$	0.200 	&	452.302 	&	0.60 	&	9.10 	&	2.310 	&	2.700 	&	$	54.20 	$	&	$	-63.70 	$	&	538.4	\\
FRB20190422A	&	2	&	48.56	$\pm$	0.200 	&	35.15	$\pm$	0.200 	&	452.302 	&	0.60 	&	9.10 	&	2.000 	&	2.700 	&	$	24.00 	$	&	$	-32.00 	$	&	417.5	\\
FRB20190423B	&	0	&	298.58	$\pm$	0.210 	&	26.19	$\pm$	0.210 	&	584.949 	&	0.87 	&	7.00 	&	2.490 	&	3.000 	&	$	62.40 	$	&	$	-106.00 	$	&	400.2	\\
FRB20190423B	&	1	&	298.58	$\pm$	0.210 	&	26.19	$\pm$	0.210 	&	584.949 	&	0.87 	&	7.00 	&	8.500 	&	3.000 	&	$	63.00 	$	&	$	-116.00 	$	&	400.2	\\
FRB20190425B	&	0	&	210.12	$\pm$	0.100 	&	88.6	$\pm$	0.210 	&	1031.724 	&	1.25 	&	3.10 	&	1.108 	&	1.300 	&	$	22.40 	$	&	$	-65.60 	$	&	400.2	\\
FRB20190426A	&	0	&	115.04	$\pm$	0.200 	&	59.12	$\pm$	0.200 	&	340.662 	&	1.59 	&	2.01 	&	0.398 	&	0.430 	&	$	27.00 	$	&	$	-71.70 	$	&	400.2	\\
FRB20190428A	&	0	&	170.73	$\pm$	0.027 	&	23.33	$\pm$	0.150 	&	969.400 	&	2.22 	&	7.40 	&	0.374 	&	3.630 	&	$	54.00 	$	&	$	-48.80 	$	&	400.2	\\
FRB20190429B	&	0	&	329.93	$\pm$	0.240 	&	3.96	$\pm$	0.330 	&	295.650 	&	0.74 	&	5.00 	&	6.380 	&	7.800 	&	$	99.00 	$	&	$	-910.00 	$	&	400.2	\\
FRB20190430A	&	0	&	77.7	$\pm$	0.240 	&	87.01	$\pm$	0.320 	&	339.250 	&	0.75 	&	7.70 	&	3.380 	&	3.230 	&	$	4.70 	$	&	$	-29.10 	$	&	400.2	\\
FRB20190502C	&	0	&	155.6	$\pm$	0.150 	&	82.97	$\pm$	0.040 	&	396.835 	&	3.60 	&	8.30 	&	0.527 	&	0.320 	&	$	9.00 	$	&	$	-26.80 	$	&	400.2	\\
FRB20190515D	&	0	&	67.13	$\pm$	0.210 	&	-5.01	$\pm$	0.340 	&	426.061 	&	3.00 	&	8.80 	&	0.420 	&	1.490 	&	$	30.60 	$	&	$	-54.30 	$	&	400.2	\\
FRB20190517C	&	0	&	87.5	$\pm$	0.190 	&	26.62	$\pm$	0.200 	&	335.575 	&	3.10 	&	8.70 	&	0.377 	&	0.149 	&	$	8.48 	$	&	$	-50.20 	$	&	400.2	\\
FRB20190518G	&	0	&	94.79	$\pm$	0.200 	&	75.52	$\pm$	0.140 	&	524.946 	&	0.99 	&	1.76 	&	0.666 	&	0.720 	&	$	19.80 	$	&	$	-75.30 	$	&	400.2	\\
FRB20190519E	&	0	&	168.28	$\pm$	0.200 	&	41.65	$\pm$	0.200 	&	693.830 	&	1.00 	&	1.46 	&	0.414 	&	0.450 	&	$	2.00 	$	&	$	4.10 	$	&	400.2	\\
FRB20190519F	&	0	&	165.63	$\pm$	0.200 	&	77.23	$\pm$	0.220 	&	797.766 	&	0.75 	&	4.00 	&	1.360 	&	1.360 	&	$	21.70 	$	&	$	-86.40 	$	&	400.2	\\
FRB20190519J	&	0	&	296.21	$\pm$	0.085 	&	86.93	$\pm$	0.300 	&	642.759 	&	0.63 	&	1.70 	&	0.460 	&	0.501 	&	$	24.30 	$	&	$	-259.00 	$	&	400.2	\\
FRB20190520A	&	0	&	273.52	$\pm$	0.026 	&	26.32	$\pm$	0.094 	&	432.506 	&	1.08 	&	2.40 	&	0.649 	&	0.710 	&	$	11.40 	$	&	$	-54.30 	$	&	400.2	\\
FRB20190527A	&	0	&	12.45	$\pm$	0.200 	&	7.99	$\pm$	0.067 	&	584.580 	&	0.47 	&	10.10 	&	2.670 	&	5.080 	&	$	47.00 	$	&	$	-122.00 	$	&	540.6	\\
FRB20190527A	&	1	&	12.45	$\pm$	0.200 	&	7.99	$\pm$	0.067 	&	584.580 	&	0.47 	&	10.10 	&	2.470 	&	5.080 	&	$	30.70 	$	&	$	-133.00 	$	&	400.2	\\
FRB20190529A	&	0	&	68.06	$\pm$	0.220 	&	40.32	$\pm$	0.230 	&	704.450 	&	0.47 	&	1.45 	&	1.040 	&	1.500 	&	$	24.20 	$	&	$	-97.00 	$	&	441.8	\\
FRB20190530A	&	0	&	68.74	$\pm$	0.025 	&	60.59	$\pm$	0.200 	&	555.445 	&	0.58 	&	1.69 	&	1.020 	&	1.300 	&	$	17.70 	$	&	$	-91.00 	$	&	400.2	\\
FRB20190531C	&	0	&	331.14	$\pm$	0.240 	&	43	$\pm$	0.240 	&	478.202 	&	0.37 	&	1.20 	&	1.450 	&	1.900 	&	$	18.30 	$	&	$	-74.00 	$	&	400.2	\\
FRB20190531E	&	0	&	15.2	$\pm$	0.260 	&	0.54	$\pm$	0.370 	&	328.203 	&	2.70 	&	5.30 	&	0.465 	&	0.560 	&	$	1.10 	$	&	$	4.60 	$	&	400.2	\\
FRB20190601B	&	0	&	17.88	$\pm$	0.170 	&	23.82	$\pm$	0.036 	&	787.795 	&	1.00 	&	13.00 	&	4.040 	&	5.670 	&	$	9.70 	$	&	$	-68.60 	$	&	400.2	\\
FRB20190601C	&	0	&	88.52	$\pm$	0.180 	&	28.47	$\pm$	0.057 	&	424.066 	&	1.32 	&	5.80 	&	0.684 	&	0.119 	&	$	35.30 	$	&	$	-68.80 	$	&	400.2	\\
FRB20190601C	&	1	&	88.52	$\pm$	0.180 	&	28.47	$\pm$	0.057 	&	424.066 	&	1.32 	&	5.80 	&	0.510 	&	0.119 	&	$	36.10 	$	&	$	-79.50 	$	&	400.2	\\
FRB20190605C	&	0	&	168.32	$\pm$	0.048 	&	-5.19	$\pm$	0.093 	&	187.642 	&	4.60 	&	4.40 	&	0.495 	&	0.520 	&	$	9.90 	$	&	$	-40.50 	$	&	400.2	\\
FRB20190605D	&	0	&	26.72	$\pm$	0.230 	&	28.62	$\pm$	0.250 	&	1656.533 	&	0.82 	&	2.16 	&	1.069 	&	1.200 	&	$	48.30 	$	&	$	-50.80 	$	&	400.2	\\
FRB20190609A	&	0	&	345.3	$\pm$	0.280 	&	87.94	$\pm$	0.330 	&	316.643 	&	3.60 	&	10.40 	&	0.432 	&	0.500 	&	$	62.40 	$	&	$	-84.00 	$	&	400.2	\\
FRB20190609A	&	1	&	345.3	$\pm$	0.280 	&	87.94	$\pm$	0.330 	&	316.643 	&	3.60 	&	10.40 	&	2.120 	&	0.500 	&	$	55.00 	$	&	$	-67.00 	$	&	400.2	\\
FRB20190612B	&	0	&	222.21	$\pm$	0.220 	&	4.31	$\pm$	0.120 	&	187.600 	&	2.41 	&	3.78 	&	0.186 	&	0.121 	&	$	10.60 	$	&	$	-25.90 	$	&	400.2	\\
FRB20190614A	&	0	&	179.79	$\pm$	0.091 	&	88.33	$\pm$	0.240 	&	1064.039 	&	0.83 	&	2.21 	&	1.300 	&	0.770 	&	$	1.60 	$	&	$	-29.00 	$	&	400.2	\\
FRB20190617B	&	0	&	56.43	$\pm$	0.020 	&	1.16	$\pm$	0.270 	&	273.510 	&	0.99 	&	9.20 	&	7.580 	&	9.500 	&	$	10.60 	$	&	$	-38.40 	$	&	400.2	\\
FRB20190617C	&	0	&	134.37	$\pm$	0.210 	&	35.7	$\pm$	0.210 	&	640.156 	&	0.54 	&	4.10 	&	3.880 	&	4.700 	&	$	10.40 	$	&	$	-8.60 	$	&	400.2	\\
FRB20190618A	&	0	&	321.25	$\pm$	0.170 	&	25.44	$\pm$	0.075 	&	228.947 	&	2.40 	&	4.30 	&	0.548 	&	0.580 	&	$	3.34 	$	&	$	-35.80 	$	&	400.2	\\
FRB20190621C	&	0	&	206.57	$\pm$	0.210 	&	5.23	$\pm$	0.290 	&	570.267 	&	1.98 	&	2.38 	&	0.443 	&	0.510 	&	$	39.10 	$	&	$	-101.00 	$	&	404.7	\\
FRB20190621D	&	0	&	270.63	$\pm$	0.016 	&	78.89	$\pm$	0.170 	&	647.514 	&	0.89 	&	4.30 	&	2.290 	&	2.600 	&	$	7.40 	$	&	$	-22.30 	$	&	400.2	\\
FRB20190623B	&	0	&	335.22	$\pm$	0.180 	&	46.12	$\pm$	0.180 	&	1556.765 	&	1.58 	&	2.78 	&	0.440 	&	0.720 	&	$	54.20 	$	&	$	-57.10 	$	&	400.2	\\
FRB20190624B	&	0	&	304.65	$\pm$	0.068 	&	73.61	$\pm$	0.200 	&	213.922 	&	16.50 	&	20.00 	&	0.372 	&	0.400 	&	$	34.80 	$	&	$	-43.20 	$	&	400.2	\\
FRB20190625D	&	0	&	115.02	$\pm$	0.013 	&	4.87	$\pm$	0.033 	&	717.883 	&	5.30 	&	12.10 	&	0.687 	&	0.720 	&	$	17.84 	$	&	$	-76.10 	$	&	400.2	\\
FRB20190627A	&	0	&	195.89	$\pm$	0.240 	&	0.75	$\pm$	0.370 	&	404.221 	&	1.98 	&	2.62 	&	0.663 	&	0.800 	&	$	9.30 	$	&	$	-27.30 	$	&	400.2	\\
FRB20190629A	&	0	&	6.34	$\pm$	0.250 	&	12.67	$\pm$	0.260 	&	503.779 	&	0.82 	&	3.05 	&	1.140 	&	1.700 	&	$	24.70 	$	&	$	-35.30 	$	&	400.2	\\
FRB20190701C	&	0	&	96.36	$\pm$	0.230 	&	81.63	$\pm$	0.270 	&	974.195 	&	0.88 	&	2.50 	&	1.440 	&	1.800 	&	$	46.20 	$	&	$	-211.00 	$	&	400.2	\\
FRB20190701D	&	0	&	112.1	$\pm$	0.180 	&	66.7	$\pm$	0.160 	&	933.363 	&	1.33 	&	8.60 	&	1.400 	&	1.530 	&	$	6.49 	$	&	$	-20.90 	$	&	400.2	\\
\enddata
\end{deluxetable*}

\end{document}